\documentclass[journal]{IEEEtran}

\input{preamble}

\begin{document}

\title{Breaking the Multi-Enhancement Bottleneck: Domain-Consistent Quality Enhancement for Compressed Images}

\author{Qunliang~Xing,~\IEEEmembership{Graduate~Student~Member,~IEEE,}
Mai~Xu,~\IEEEmembership{Senior~Member,~IEEE,}
Jing~Yang,
and~Shengxi~Li,~\IEEEmembership{Member,~IEEE}%
\thanks{Qunliang Xing, Mai Xu, Jing Yang, and Shengxi Li are with the School of Electronic Information Engineering, Beihang University, Beijing 100191, China. Qunliang Xing is also with the Shen Yuan Honors College, Beihang University, Beijing 100191, China (e-mail: ryanxingql@gmail.com; maixu@buaa.edu.cn; jing\_yang@buaa.edu.cn; lishengxi@buaa.edu.cn). \textit{(Corresponding author: Mai Xu.)}}%
}

\markboth{Breaking the Multi-Enhancement Bottleneck: Domain-Consistent Quality Enhancement for Compressed Images}
{XING~\MakeLowercase{et~al.}: Breaking the Multi-Enhancement Bottleneck: Domain-Consistent Quality Enhancement for Compressed Images}

\maketitle

\begin{abstract}
    Quality enhancement methods have been widely integrated into visual communication pipelines to mitigate artifacts in compressed images. Ideally, these quality enhancement methods should perform robustly when applied to images that have already undergone prior enhancement during transmission. We refer to this scenario as multi-enhancement, which generalizes the well-known multi-generation scenario of image compression. Unfortunately, current quality enhancement methods suffer from severe degradation when applied in multi-enhancement. To address this challenge, we propose a novel adaptation method that transforms existing quality enhancement models into domain-consistent ones. Specifically, our method enhances a low-quality compressed image into a high-quality image within the natural domain during the first enhancement, and ensures that subsequent enhancements preserve this quality without further degradation. Extensive experiments validate the effectiveness of our method and show that various existing models can be successfully adapted to maintain both fidelity and perceptual quality in multi-enhancement scenarios.
\end{abstract}
    
\begin{IEEEkeywords}
    Multi-enhancement, quality enhancement, compressed images, domain-consistent.
\end{IEEEkeywords}

\section{Introduction}

\IEEEPARstart{W}{e} are embracing an era of explosive growth in visual data.
According to the Domo statistics~\cite{domoDataNeverSleeps2023}, internet users watched the equivalent of 43 years' worth of streaming content every minute in 2023.
The rapid advancement of Artificial Intelligence (AI) has further accelerated this growth, generating approximately 34 million AI-generated images per day in 2024~\cite{everypixeljournalAIImageStatistics2023}.
To store and transmit this vast amount of data, lossy yet highly efficient image compression standards---such as Joint Photographic Experts Group (JPEG)~\cite{wallaceJPEGStillPicture1992}, JPEG 2000~\cite{marcellinOverviewJPEG20002000}, and High Efficiency Video Coding with Main Still Picture profile (HEVC-MSP)~\cite{sullivanOverviewHighEfficiency2012}---have been developed to reduce bandwidth and storage requirements.
However, compressed images often suffer from artifacts such as blocking, ringing, and blurring, which severely degrade end-user Quality of Experience (QoE)~\cite{seshadrinathanStudySubjectiveObjective2010,itu-tVocabularyPerformanceQuality2017} and impair the performance of downstream applications~\cite{duboisLossyCompressionLossless2021,ehrlichAnalyzingMitigatingJPEG2021}.

To address this challenge, quality enhancement methods have been widely integrated into visual communication pipelines to reduce visual artifacts in compressed images.
For example, the HEVC compression pipeline incorporates in-loop deblocking~\cite{norkinHEVCDeblockingFilter2012} and Sample Adaptive Offset (SAO)~\cite{fuSampleAdaptiveOffset2012} filters to enhance the quality of intra-codec compressed images.
In addition, recent compression standards such as JPEG AI~\cite{iso/iecjtc1/sc29/wg1WhitePaperJPEG2021} and Versatile Video Coding (VVC)~\cite{brossOverviewVersatileVideo2021} have also adopted quality enhancement methods to improve visual quality of compressed images.
Furthermore, we have witnessed rapid growth in quality enhancement methods arising from both industrial applications (e.g., digital cameras, mobile devices, streaming servers, televisions, monitors, and virtual reality systems) and academic research~\cite{wangNovelDeepLearningbased2017,yangMultiframeQualityEnhancement2018,xingEarlyExitNot2020a,dengSpatioTemporalDeformableConvolution2020,guanMFQE20New2021,yangNTIRE2022Challenge2022,zhengProgressiveTrainingTwoStage2022,xingDAQEEnhancingQuality2023,ehrlichLeveragingBitstreamMetadata2024,xingEnhancingQualityCompressed2024}.

\begin{figure*}[!t]
    \refstepcounter{figure}
    \def\currentfigurelabel{fig:multi-enhancement}
    \centering
    \begin{overpic}[trim={0em 1em 0em 0em}, clip, width=\linewidth]{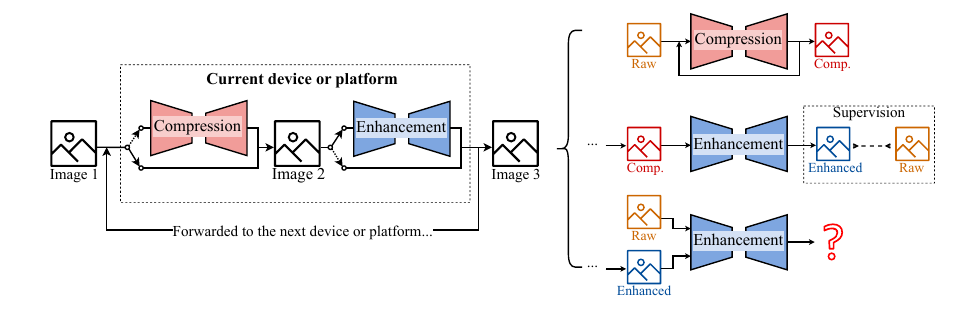}
        \putwithlabel{29}{4}{(a)}
        \putwithlabel{74.8}{21}{(b)}
        \putwithlabel{74.8}{11}{(c)}
        \putwithlabel{74.8}{1}{(d)}
    \end{overpic}
    \addtocounter{figure}{-1}%
    \caption{Motivation for our domain-consistent quality enhancement method.
    (a) Prevalent multi-enhancement scenario, where devices or platforms may include or omit compression or enhancement processes.
    (b) Well-studied multi-generation scenario in image compression~\cite{sorialMultigenerationTransformcodedImages1997}.
    (c) Typical scenario where existing enhancement methods effectively process compressed images.
    (d) Challenging scenario where enhancement methods must handle high-quality images already in the natural domain.}
    \label{fig:multi-enhancement}
\end{figure*}

\begin{figure}[!t]
    \centering
    \includegraphics[trim={2em 1em 2em 1em}, clip, width=\linewidth]{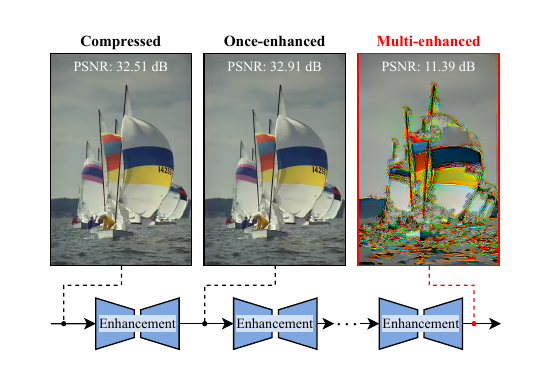}
    \caption{Subjective degradation in the multi-enhancement scenario. The raw image \textit{kodim09}~\cite{kodakKodakLosslessTrue1999} is compressed using Better Portable Graphics (BPG)~\cite{bellardBetterPortableGraphics2018} with a Quality Factor (QF) of 37. Note that BPG implements the HEVC-MSP image compression standard. The compressed image is then enhanced by a classic quality enhancement method: the Artifact Reduction Convolutional Neural Network (AR-CNN)~\cite{dongCompressionArtifactsReduction2015}.}
    \label{fig:multi-enhancement-demo}
\end{figure}

However, the widespread adoption of quality enhancement methods has introduced a new challenge.
In real-world scenarios, the input is often an already enhanced image, rather than a raw compressed image without enhancement.
Such cases are common in visual communication pipelines, where a compressed image is enhanced by various devices or platforms, and then further enhanced downstream.
We refer to this scenario as \textbf{multi-enhancement}, as illustrated in \cref{fig:multi-enhancement:a}, which generalizes the well-studied multi-generation scenario of image compression~\cite{sorialMultigenerationTransformcodedImages1997}.
A brief introduction to the multi-generation scenario is provided in \cref{sec:related}.
In the multi-enhancement scenario, the enhancement process may receive either a low-quality image (i.e., a compressed image) or a high-quality image (i.e., an already enhanced image or a raw image without compression).
Current quality enhancement methods are effective on compressed images, as they are designed to map images from the compression domain to the natural domain.
However, we observe significant quality degradation when these methods are applied to high-quality images, as shown in \cref{fig:multi-enhancement-demo}.
This degradation arises because both enhanced and raw images already reside in the natural domain, while existing enhancement methods are not optimized for such high-quality inputs.

In this paper, we address the multi-enhancement challenge by introducing a domain-consistent quality enhancement method.
We formulate this challenge as an idempotent enhancement problem and propose a novel adaptation method that transforms existing quality enhancement models into domain-consistent ones.
Our method enhances a low-quality compressed image from the compression domain into a high-quality image in the natural domain during the first enhancement and ensures that the image remains in the natural domain during subsequent enhancements without significant quality degradation.
Extensive experiments validate the effectiveness of the proposed method and demonstrate that various existing quality enhancement models can be successfully adapted to maintain both high fidelity and perceptual quality in multi-enhancement scenarios.

To summarize, the contributions of this paper are as follows:
\begin{itemize}
    \item {We generalize the traditional multi-generation scenario of image compression into the modern and prevalent multi-enhancement scenario. Based on this, we reveal that existing quality enhancement methods suffer from severe degradation in multi-enhancement.}
    \item {We propose a novel adaptation method that transforms existing quality enhancement models into domain-consistent ones. These adapted models effectively enhance image quality while preserving domain consistency of the enhanced image across successive enhancements.}
    \item {We conduct extensive experiments to validate the effectiveness of the proposed method, demonstrating that various existing quality enhancement models can be successfully adapted to maintain both high fidelity and perceptual quality in multi-enhancement scenarios.}
\end{itemize}

\section{Related Work}
\label{sec:related}

\subsection{Quality Enhancement for Compressed Images}

Over the past decade, extensive research has been devoted to enhancing the quality of compressed images.
One category is intra-codec enhancement, where the enhancement is integrated within the image compression process itself.
For example, the HEVC compression standard~\cite{sullivanOverviewHighEfficiency2012} incorporates deblocking~\cite{norkinHEVCDeblockingFilter2012} and SAO~\cite{fuSampleAdaptiveOffset2012} filters to suppress the blocking effects in intra-codec compressed images.
In this paradigm, enhancement is embedded as part of the compression process and is typically studied under the multi-generation scenario, as illustrated by \cref{fig:multi-enhancement:b}.

Meanwhile, greater attention has been given to post-codec enhancement, also known as decoder-side enhancement~\cite{shenReviewPostprocessingTechniques1998,dingAdvancesVideoCompression2021}, as illustrated in \cref{fig:multi-enhancement:c}.
This approach can be easily integrated into visual communication pipelines without altering the codec itself.
A pioneering example is the shallow four-layer AR-CNN proposed by Dong~et~al.~\cite{dongCompressionArtifactsReduction2015}, which is inspired by dictionary learning and includes four stages: feature extraction, enhancement, mapping, and reconstruction.
Subsequently, the Denoising CNN (DnCNN)~\cite{zhangGaussianDenoiserResidual2017} introduced a much deeper 20-layer network to suppress JPEG artifacts, leveraging residual learning~\cite{heDeepResidualLearning2016} and Batch Normalization (BN)\cite{ioffeBatchNormalizationAccelerating2015} to facilitate training.
At the same time, Wang~et~al.~\cite{wangNovelDeepLearningbased2017} introduced the Deep CNN-based Auto Decoder (DCAD) for decoder-side quality enhancement.
Notably, these methods are not tailored to compression quality enhancement and thus function as general-purpose solutions applicable to many image enhancement tasks, such as image denoising.

Additionally, several methods explicitly leverage the properties of compressed images.
For example, the Deep Dual-Domain (D3) method~\cite{wangD3DeepDualdomain2016} and the Deep Dual-domain Convolutional neural Network (DDCN)~\cite{guoBuildingDualdomainRepresentations2016} utilize JPEG quantization priors to mitigate JPEG compression artifacts.
The Resource-efficient Blind Quality Enhancement (RBQE) method~\cite{xingEarlyExitNot2020a} enables blind quality enhancement with progressive enhancement and early termination once acceptable quality is achieved.
Ehrlich~et~al.~\cite{ehrlichQuantizationGuidedJPEG2020} proposed enhancing DCT coefficients of JPEG images using the JPEG quantization matrix.
Jiang~et~al.~\cite{jiangFlexibleBlindJPEG2021} introduced a blind enhancement method that predicts quantization factors (QF).
To further improve efficiency, the Defocus-Aware Quality Enhancement (DAQE) method~\cite{xingDAQEEnhancingQuality2023} performs region-wise divide-and-conquer enhancement based on defocus estimation.

Unfortunately, existing methods struggle with high-quality inputs---such as already enhanced or raw images---which are common in practice due to multiple enhancement operations across diverse devices and platforms.
To address this limitation, this paper introduces a novel domain-consistent quality enhancement method designed to handle such scenarios.

\subsection{Multi-Generation Robustness of Image Compression}

In this paper, we generalize the traditional multi-generation scenario~\cite{sorialMultigenerationTransformcodedImages1997} in image compression to a more modern and prevalent scenario: the multi-enhancement concerning quality enhancement.
To support this generalization, we briefly review the concept of multi-generation robustness in image compression.
In many applications---such as image editing, distribution, and post-production (e.g., visual effects creation)---an image may undergo multiple rounds of re-compression, a process known as multi-generation.
Robustness to such re-compression is a critical property of compression standards like JPEG, Moving Picture Experts Group (MPEG)~\cite{legallMPEGVideoCompression1991}, JPEG 2000, Advanced Video Coding (AVC)~\cite{wiegandOverviewH264AVC2003}, and HEVC.
This multi-generation robustness has been extensively studied in academic research~\cite{erdemMultigenerationCharacteristicsMPEG1994,horneStudyCharacteristicsMPEG21996,joshiComparisonMultipleCompression2000,linAchievingReLossFreeVideo2009,zhuIdempotentH264Intraframe2009,stankowskiVideoQualityMultiple2013}, and more recently for learning-based compression methods~\cite{kimSuccessiveLearnedImage2022,liIdempotentLearnedImage2023,liImprovingMultigenerationRobustness2023,xu2024idempotence}.

\begin{table*}[!t]
    \caption{Degradation Index (DI) values indicating quality degradation speed under multi-enhancement}
    \label{tab:quality-degradation}
    
    \centering
    \begin{tabular}{c c c c c c c c c c c c c c c c c}
        \toprule
        \multirow{2}{*}{Dataset} & \multirow{2}{*}{Codec} & \multirow{2}{*}{QP/QF} & \multicolumn{11}{c}{DI (\%) in terms of PSNR} & SSIM & TOP. & LPIPS\\
        \cmidrule{4-17}
         & & & \cite{dongCompressionArtifactsReduction2015} & \cite{zhangGaussianDenoiserResidual2017} & \cite{wangNovelDeepLearningbased2017} & \cite{zhangResidualDenseNetwork2018} & \cite{guoConvolutionalBlindDenoising2019} & \cite{xingEarlyExitNot2020a} & \cite{zamirMultistageProgressiveImage2021} & \cite{ledigPhotoRealisticSingleImage2017} & \cite{wangESRGANEnhancedSuperResolution2018} & \cite{wangRealESRGANTrainingRealWorld2021} & Ave. & Ave. & Ave. & Ave.\\
        \midrule
        \multirow{5}{*}{DIV2K} & \multirow{5}{*}{BPG} & 27 & 2.11 & 3.86 & 3.72 & 3.50 & 4.17 & 4.23 & 4.60 & 8.40 & 8.53 & 7.73 & 5.08 & 3.36 & 5.19 & 61.48\\
         & & 32 & 1.77 & 2.71 & 2.35 & 2.45 & 3.50 & 3.57 & 3.22 & 8.83 & 9.72 & 6.52 & 4.46 & 3.68 & 6.30 & 40.67\\
         & & 37 & 1.71 & 2.09 & 2.06 & 2.22 & 3.19 & 2.71 & 2.64 & 8.76 & 9.98 & 6.41 & 4.18 & 4.11 & 6.80 & 25.94\\
         & & 42 & 1.82 & 2.04 & 2.04 & 2.14 & 3.13 & 2.50 & 2.59 & 7.70 & 8.77 & 6.86 & 3.96 & 3.88 & 6.06 & 14.81\\
        & & Ave. & 1.85 & 2.67 & 2.54 & 2.58 & 3.50 & 3.25 & 3.26 & 8.42 & 9.25 & 6.88 & 4.42 & 3.76 & 6.09 & 35.73\\
        \midrule
        DIV2K & JPEG & Ave. & 2.37 & 2.83 & 2.92 & 3.02 & 3.51 & 3.48 & 3.39 & 6.90 & 7.04 & 5.51 & 4.10 & 2.88 & 6.80 & 53.95\\
        \midrule
        Flickr2K & BPG & Ave. & 2.06 & 2.97 & 2.81 & 2.87 & 3.92 & 3.68 & 3.75 & 8.79 & 9.49 & 7.26 & 4.76 & 3.80 & 6.50 & 39.96\\
        \midrule
        Flickr2K & JPEG & Ave. & 2.54 & 3.05 & 3.17 & 3.25 & 4.02 & 3.84 & 3.89 & 7.38 & 7.57 & 5.95 & 4.47 & 2.92 & 7.27 & 56.45\\
        \bottomrule
    \end{tabular}
\end{table*}

However, we observe severe quality degradation in prevalent visual communication pipelines.
The core issue is that existing notions of multi-generation robustness overlook the widespread presence of quality enhancement processes throughout these pipelines.
We refer to the corresponding requirement as multi-enhancement robustness, which is the central focus of this paper.
Accordingly, in the following section, we first investigate the quality degradation that occurs under multi-enhancement, and then propose our domain-consistent quality enhancement method to address it.

\section{Challenges of Multi-Enhancement}
\label{sec:findings}

This section investigates the quality degradation that arises in the multi-enhancement scenario, particularly as illustrated in \cref{fig:multi-enhancement:d}.

\begin{observation}
    Multi-enhancement leads to severe quality degradation for existing methods, with PSNR dropping by over 4\% per enhancement cycle on average.
    \label{ob:1}
\end{observation}

\textit{Analysis:}
We conduct our investigation using two widely adopted high-quality image datasets: DIVerse 2K (DIV2K)~\cite{agustssonNTIRE2017Challenge2017} and Flickr2K~\cite{limEnhancedDeepResidual2017}.
All images are compressed using the commonly used JPEG codec and the more efficient Better Portable Graphics (BPG) codec~\cite{bellardBetterPortableGraphics2018}, which implements the HEVC-MSP standard.
Each codec is applied with four compression levels: JPEG uses QF of 30/40/50/60, and BPG uses Quantization Parameters (QP) of 27/32/37/42.
The compressed images are then enhanced using several commonly used quality enhancement baselines, including AR-CNN, DnCNN, DCAD, and RBQE.
We further evaluate more advanced methods such as the Residual Dense Network (RDN)~\cite{zhangResidualDenseNetwork2018}, the Convolutional Blind Denoising Network (CBDNet)~\cite{guoConvolutionalBlindDenoising2019}, and the Multi-stage Progressive image Restoration Network (MPRNet)~\cite{zamirMultistageProgressiveImage2021}.
Additionally, we include methods based on Generative Adversarial Networks (GANs)~\cite{goodfellowGenerativeAdversarialNets2014}, which are widely used for perceptual enhancement: the Super-Resolution GAN (SRGAN)~\cite{ledigPhotoRealisticSingleImage2017}, the Enhanced SRGAN (ESRGAN)~\cite{wangESRGANEnhancedSuperResolution2018}, and the Real-ESRGAN~\cite{wangRealESRGANTrainingRealWorld2021}.
All methods are trained on the DIV2K training set and evaluated on the DIV2K validation set and the last 100 images of the Flickr2K dataset.\footnote{All methods are available at \url{https://github.com/ryanxingql/powerqe}.}
During testing, each compressed image undergoes five enhancement cycles using the same method.

\begin{figure*}[!t]
    \centering
    \includegraphics[trim={0em 0em 0em 0em}, clip, width=1\linewidth]{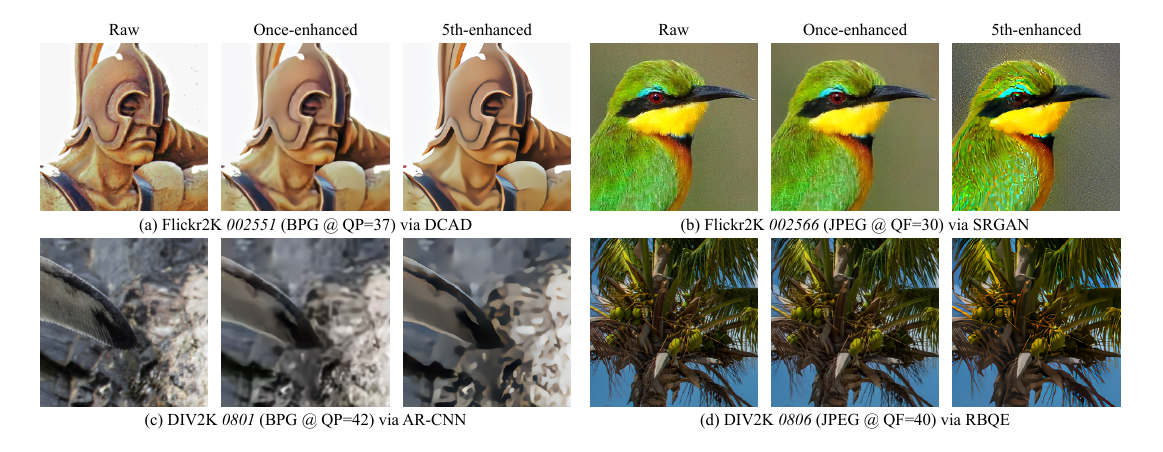}
    \caption{Subjective results illustrating quality degradation under multi-enhancement across various datasets, codecs, compression levels, and enhancement methods.}
    \label{fig:quality-degradation-subjective}
\end{figure*}

\begin{figure}[!t]
    \centering
    \includegraphics[trim={0em 0em 0em 0em}, clip, width=\linewidth]{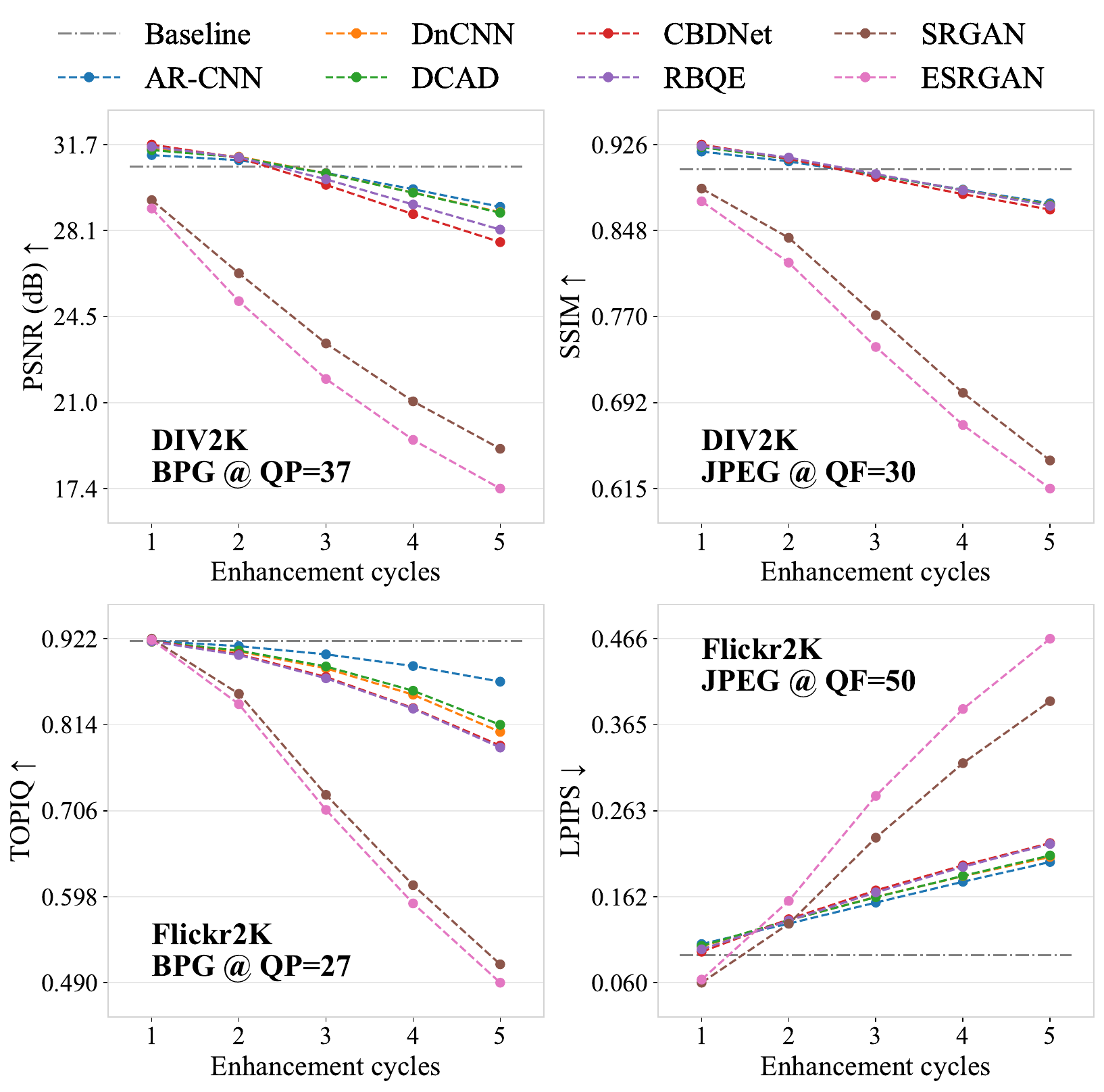}
    \caption{Objective results illustrating quality degradation under multi-enhancement across various datasets, codecs, compression levels, and evaluation metrics.}
    \label{fig:quality-degradation}
\end{figure}

We evaluate the quality of enhanced images using both fidelity and perceptual quality metrics.
For fidelity, we use the Peak Signal-to-Noise Ratio (PSNR) and the Structural SIMilarity (SSIM)~\cite{wangImageQualityAssessment2004}.
For perceptual quality, we adopt the TOP-down Image Quality (TOPIQ)~\cite{chenTOPIQTopDownApproach2024} and the Learned Perceptual Image Patch Similarity (LPIPS)~\cite{zhangUnreasonableEffectivenessDeep2018}.
\Cref{fig:quality-degradation} presents representative results across different datasets, codecs, compression levels, and evaluation metrics.
From these observations, we draw three main conclusions:
\begin{itemize}
    \item [(1)]{GAN-based methods suffer from more severe quality degradation than CNN-based methods in terms of both fidelity and perceptual metrics.}
    \item [(2)]{Image quality deteriorates significantly in the multi-enhancement scenario, always falling below the baseline compression quality after only one or two enhancement cycles.}
    \item [(3)]{Stronger baseline models tend to exhibit weaker robustness under multi-enhancement. For instance, although CBDNet achieves higher PSNR quality than AR-CNN in the initial enhancement, it degrades more rapidly in subsequent enhancements.}
\end{itemize}

Furthermore, we quantify the degradation speed using a proposed metric called the Degradation Index (DI).
Given the near-linear trends observed in \cref{fig:quality-degradation}, we define DI as the degradation slope as follows:
\begin{equation}
    \text{DI} = m \cdot \frac{(Q_1 - Q_n) / Q_1}{n - 1} \times 100\%,
\end{equation}
where \( n \) is the number of enhancement cycles, \( Q_1 \) and \( Q_n \) are the quality values after the first and the \(n\)-th enhancement, respectively, and \( m = 1 \) if higher quality values indicate better performance (e.g., PSNR), or \( m = -1 \) otherwise (e.g., LPIPS).  
The normalization by \( Q_1 \) ensures that DI is independent of absolute quality levels.

As shown in \cref{tab:quality-degradation}, fidelity-based metrics degrade by up to 5.08\% and 4.11\% per enhancement cycle in terms of PSNR and SSIM, respectively.
Perceptual quality metrics exhibit even more severe degradation, reaching up to 7.27\% for TOPIQ and 61.48\% for LPIPS per cycle.
These results confirm that current methods suffer from significant performance drops under multi-enhancement.

\begin{observation}
    Existing methods tend to generate comic-style or oil-painting-like artifacts in the multi-enhancement scenario.
    \label{ob:2}
\end{observation}

\textit{Analysis:}
\Cref{fig:quality-degradation-subjective} presents subjective results illustrating quality degradation under multi-enhancement across various datasets, codecs, compression levels, and enhancement methods.
As shown, multi-enhanced images degrade significantly in visual quality compared to both the raw image and the once-enhanced image.
In particular, the fifth-enhanced images exhibit prominent visual artifacts, including distorted colors, over-smoothed edges, and unnatural textures.
These artifacts can be characterized as comic-style or oil-painting-like.
When the number of enhancement cycles increases further, such artifacts may even corrupt the original image content, as illustrated in \cref{fig:multi-enhancement-demo}.

\begin{figure}[!t]
    \centering
    \includegraphics[trim={0em 0em 0em 0em}, clip, width=1\linewidth]{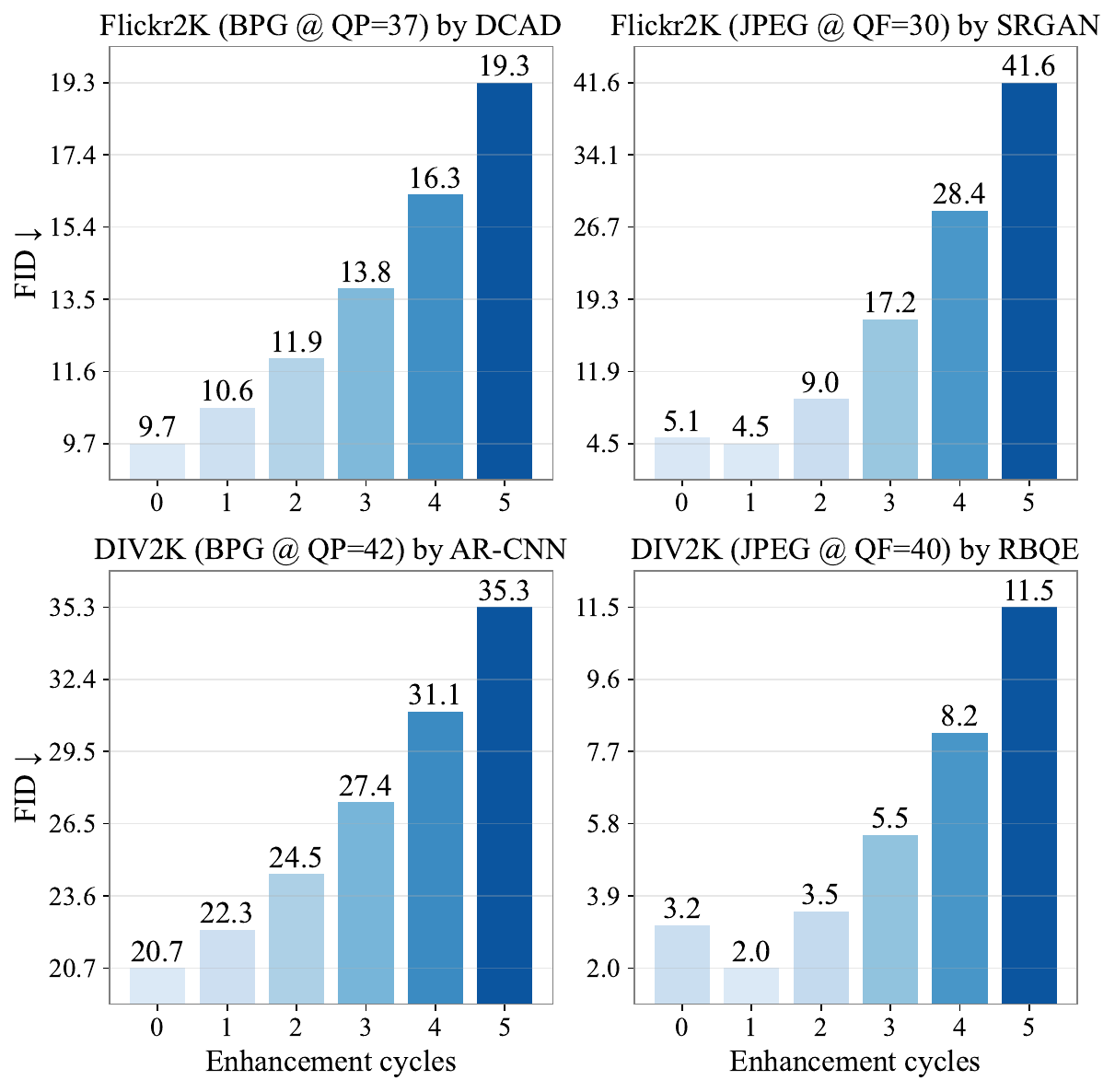}
    \caption{FID results illustrating quality degradation under multi-enhancement across various datasets, codecs, compression levels, and enhancement methods.}
    \label{fig:quality-degradation-fid}
\end{figure}

To the best of our knowledge, no existing work has studied these comic-style or oil-painting-like artifacts.
Intuitively, they enlarge the domain gap between the enhancement domain and natural domain, where enhanced and raw images reside, respectively.
To quantify this domain gap, we adopt the Fréchet Inception Distance (FID)~\cite{heuselGANsTrainedTwo2017}.
\Cref{fig:quality-degradation-fid} shows the FID results, where Cycle-0 images represent the compressed images.
Note that smaller FID values indicate a smaller domain gap.
As shown, FID values increase significantly after the first enhancement cycle, indicating that multi-enhanced images rapidly deviate from the natural domain.
These results further highlight the visual degradation challenges in multi-enhancement.

\begin{observation}
    The multi-enhancement challenge is more severe under the naive scenario where the same method is repeatedly applied, compared to more diverse scenarios with varying methods.
    \label{ob:3}
\end{observation}

\textit{Analysis:}
In \cref{ob:1}, we evaluated multi-enhancement quality degradation in a naive setting---termed Case~1---where the same enhancement method is repeatedly applied across all cycles.  
In this section, we examine two more complex scenarios:

\begin{itemize}
    \item {In Case~2, each device applies a different enhancement method but uses the same compression setting. We simulate this by randomly selecting five methods using the same QP/QF.}
    \item {In Case~3, both enhancement methods and compression settings vary. We randomly select five methods with various QP/QF and codecs.}
\end{itemize}

\begin{table}[!t]
    \caption{DI values for each scenario, averaged over all datasets}
    \centering
    \label{tab:quality-degradation-others}

    \begin{tabular}{c c c c c}
        \toprule
        \multirow{2}{*}{Scenario} & \multicolumn{4}{c}{DI (\%) in terms of}\\
        \cmidrule{2-5}
        & PSNR & SSIM & TOPIQ & LPIPS\\
        \midrule
        Case 1 & 4.44 & 3.34 & 6.66 & 46.52\\
        Case 2 & 3.94 & 2.85 & 6.25 & 40.68\\
        Case 3 & 3.30 & 2.26 & 5.20 & 38.21\\
        \bottomrule
    \end{tabular}
\end{table}

\Cref{tab:quality-degradation-others} shows that degradation is less severe in the more diverse Cases~2 and 3, with smaller DI values compared to the naive Case~1.  
A possible explanation is that using different enhancement methods under varying compression settings helps mitigate each method's enhancement biases in the multi-enhancement process.  
By contrast, repeatedly applying the same method, as in Case~1, can exacerbate quality degradation.  
Nonetheless, the degradation remains substantial even in the more diverse cases, with an average PSNR drop exceeding 3.30\% per enhancement cycle.  
This means that an image with an initial PSNR of 30~dB after the first enhancement could lose at least 3.96~dB over the next four enhancement cycles.  
In the following section, we propose a domain-consistent quality enhancement method to address these challenges.

\section{Domain-Consistent Quality Enhancement}

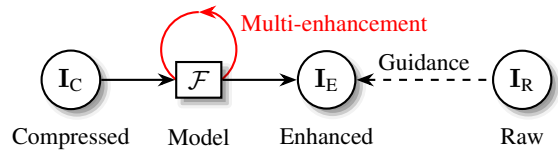
\begin{figure}[!t]
    \centering
    \begin{tikzpicture}[>=Stealth, auto, thick,
        nodeshadow/.style={circle, fill=white, blur shadow={shadow blur steps=2}, draw=black, text centered, thick, minimum size=0.75cm},
        rectshadow/.style={
            rectangle,
            draw=black,
            thick,
            fill=white,
            blur shadow={shadow blur steps=2},
            text centered,
            minimum height=0.4cm,
            minimum width=0.6cm
        }]

        \node[nodeshadow] (A) at (0,0) {\( \comp \)};
        \node[rectshadow] (B) at (1.7,0) {\( \func \)};
        \node[nodeshadow] (C) at (3.4,0) {\( \enh \)};
        \node[nodeshadow] (D) at (6,0) {\( \raw \)};

        \node at (A.south) [below=3pt] {\small Compressed};
        \node at (B.south) [below=7.5pt] {\small Model};
        \node [red] at ($(B.north)+(1.8,0.5)$) {\small Multi-enhancement};
        \node at (C.south) [below=3pt] {\small Enhanced};
        \node at (D.south) [below=3pt] {\small Raw};
      
        \draw[->] (A) -- (B);
        \draw[->] (B) -- (C);
        \draw[<-, dashed] (C) -- node[above]{\small Guidance} (D);

        \begin{pgfonlayer}{bg}
          \draw[->, thick, >=Stealth, red] ($(B.north)+(0,0.65)$) arc[start angle=90, end angle=450, radius=0.5];
        \end{pgfonlayer}
    \end{tikzpicture}
    \caption{Illustration of a straightforward solution to the multi-enhancement challenge.}
    \label{fig:straightforward}
\end{figure}

This section proposes a domain-consistent quality enhancement method to address the challenges of multi-enhancement.  
A straightforward solution is to supervise the multi-enhanced images using either the once-enhanced image or the raw image.  
This strategy has been adopted in other vision tasks, such as image compression~\cite{kimInstabilitySuccessiveDeep2020} and image deblurring~\cite{maoDeepIdempotentNetwork2023}.  
As shown in \cref{fig:straightforward}, suppose we have a high-quality raw image \( \raw \) and a low-quality compressed image \( \comp \).  
Then, these methods aim to minimize the following loss function with respect to the parameters \( \theta \) of the enhancement model \( \func \):
\begin{equation}
    \theta^{*} = \argmin_{\theta} \sum_{i=1}^{M} w_i \cdot \loss \Bigl( \func^{(i)}(\comp; \theta),\ \raw \Bigr),
\end{equation}
where \( M \) is the number of enhancement cycles (used for supervision only, not inference), \( w_i \) is the weight for the \( i \)-th cycle, and \( \loss \) is a loss function such as the Mean Squared Error (MSE) or Mean Absolute Error (MAE, also known as L1 loss).

Empirically, we find that this straightforward solution does not perform well for the quality enhancement task.  
Specifically, we re-train and evaluate the CBDNet~\cite{guoConvolutionalBlindDenoising2019} and SRGAN~\cite{ledigPhotoRealisticSingleImage2017} methods using the above loss on the DIV2K~\cite{agustssonNTIRE2017Challenge2017} dataset compressed by BPG at a QP of 37.  
We set \( M = 2 \), with weights \( w_1 = 1 \) and \( w_2 = 0.01 \), and use MAE as the loss function.  
As shown in \cref{tab:method-straightforward}, although the re-trained models exhibit slightly reduced quality degradation (as indicated by lower DI values, except for SRGAN in terms of LPIPS), their once-enhanced PSNR and LPIPS performance degrade.  
This indicates that the straightforward solution sacrifices single-enhancement quality and achieves only limited mitigation of multi-enhancement degradation.

\begin{table}[!t]
    \caption{Comparison of once-enhanced quality and DI (\%) after the fifth enhancement between baseline and straightforward solution\textsuperscript{\( \dag \)}}
    \centering
    \label{tab:method-straightforward}

    \begin{tabular}{c | c c | c c}
        \toprule
        Method & PSNR & DI & LPIPS & DI \\
        \midrule
        CBDNet & 31.66 & 3.19 & 0.21 & 5.75\\
        CBDNet\textsuperscript{\( \dag \)} & 31.56 & 2.88 & 0.24 & 4.56\\
        \midrule
        SRGAN & 29.43 & 8.76 & 0.14 & 79.22\\
        SRGAN\textsuperscript{\( \dag \)} & 29.32 & 8.72 & 0.16 & 86.24\\
        \bottomrule
    \end{tabular}
\end{table}

To address this limitation, we first model the multi-enhancement challenge as an idempotent enhancement problem and then propose a novel adaptation method that transforms existing quality enhancement methods into domain-consistent ones.  
We will show that the straightforward solution discussed above corresponds to a restricted instance of our proposed method, which explains its limited and sub-optimal performance.

\begin{figure*}[!t]
    \centering
    \includegraphics[trim={0em 0em 3em 0em}, clip, width=.8\linewidth]{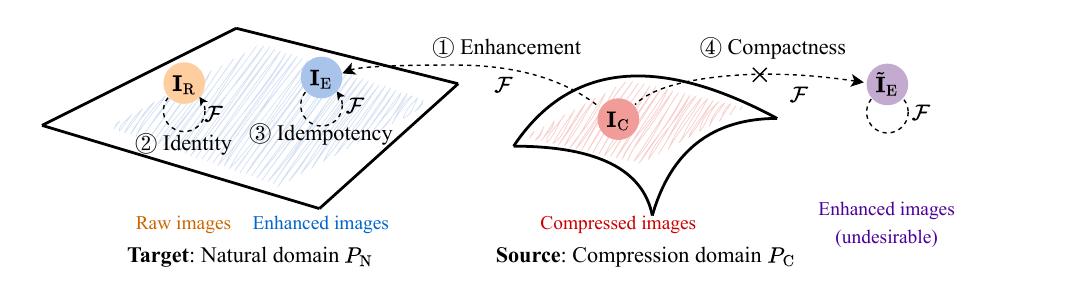}
    \caption{Illustration of two domains, four image types, and the four objectives of the proposed domain-consistent quality enhancement method.}
    \label{fig:manifold}
\end{figure*}

\subsection{Objectives for Idempotent Enhancement}

We begin by defining the notations used in this section.  
The enhancement model \( \func \) is trained to transform images from the compression domain \( \dcomp \) to the natural domain \( \dnat \).  
Here, \( \dcomp \) denotes the source domain of compressed images \( \comp \sim \dcomp \), while \( \dnat \) represents the target domain of raw images \( \raw \sim \dnat \) without compression.  
Since both types of images reside in the same dimensional space, the model \( \func \) can be applied to instances from either domain.

As mentioned earlier, we model the multi-enhancement challenge as an idempotent enhancement problem, where the goal is to learn an enhancement model \( \func \) that transforms images from the compression domain to the natural domain and preserves the enhanced images within the natural domain during subsequent enhancements.  
Specifically, this idempotent enhancement problem involves four objectives (as shown in \cref{fig:manifold}):

\subsubsection{Traditional single-enhancement objective}
The model \( \func \) is required to transform images \( \comp \) from the compression domain \( \dcomp \) to the natural domain \( \dnat \).  
For the quality enhancement task, this involves minimizing the distance between the enhanced image \( \enh \) and the corresponding ground-truth raw image \( \raw \):
\begin{equation}
    \theta^{*} = \argmin_{\theta} \lossenh ( \enh, \raw ) = \argmin_{\theta} \mathcal{D} \bigl( \func(\comp; \theta), \raw \bigr).
\end{equation}
The distance \( \mathcal{D} \) can be measured using fidelity metrics such as PSNR and SSIM~\cite{wangImageQualityAssessment2004}, perceptual metrics such as LPIPS~\cite{zhangUnreasonableEffectivenessDeep2018} and perceptual loss~\cite{johnsonPerceptualLossesRealTime2016}, or discriminators in adversarial training~\cite{goodfellowGenerativeAdversarialNets2014}.  
This objective ensures that the single-enhancement process is effective, i.e., the enhanced image \( \enh \) resides in the natural domain \( \dnat \).

\subsubsection{Identity objective}
Since the raw image \( \raw \) is already in the target natural domain \( \dnat \), the model \( \func \) should ideally behave as an identity function, i.e., \( \func(\raw) = \raw \).  
To enforce this, we minimize the distance between the raw image \( \raw \) and its enhanced version \( \func(\raw; \theta) \):
\begin{equation}
    \begin{split}
        \theta^{*} &= \argmin_{\theta} \lossiden \bigl( \func ( \raw; \theta ), \raw \bigr)\\
        &= \argmin_{\theta} \mathcal{D} \bigl( \func ( \raw; \theta ), \raw \bigr).\\
    \end{split}
\end{equation}
This objective ensures that raw images remain in the natural domain during multi-enhancement and are not unnecessarily altered by the model \( \func \).

\subsubsection{Idempotency objective}
The model \( \func \) should be idempotent---meaning that applying it multiple times should yield the same result as applying it once: \( \func \bigl( \func(\comp) \bigr) = \func(\comp) \).  
Mathematically, this can be expressed as:
\begin{equation}
    \begin{split}
        \theta^{*} &= \argmin_{\theta} \lossidem \Bigl( \func \bigl( \func(\comp; \theta); \theta \bigr), \func(\comp; \theta) \Bigr)\\
        &= \argmin_{\theta} \mathcal{D} \Bigl( \func \bigl( \func(\comp; \theta); \theta \bigr), \func(\comp; \theta) \Bigr).\\
    \end{split}
\end{equation}
This objective promotes idempotency in multi-enhancement, ensuring that repeated applications of \( \func \) do not introduce additional changes to the image.
However, this constraint alone is insufficient; thus, we introduce an additional compactness objective next.

\subsubsection{Compactness objective}
As illustrated in \cref{fig:manifold}, some enhanced images \( \unenh \) exhibit strong idempotency but low visual quality.  
These images do not reside in the target natural domain \( \dnat \) and may encourage trivial solutions for \( \func \), such as the identity function.  
To avoid such undesirable outcomes, we address this issue using the following three strategies:

\begin{itemize}
    \item [(a)] {For the idempotency objective, we focus on optimizing the initial mapping of \( \comp \), i.e., the inner instantiation of \( \func \), to restrict the idempotency domain to a sub-region of the target natural domain.  
    Mathematically, this is expressed as:
    \begin{equation}
        \begin{split}
            \theta^{*} &= \argmin_{\theta} \lossidem \Bigl( \func \bigl( \func(\comp; \theta);\hat{\theta} \bigr), \func(\comp; \theta) \Bigr)\\
            &= \argmin_{\theta} \mathcal{D} \Bigl( \func \bigl( \func(\comp; \theta); \hat{\theta} \bigr), \func(\comp; \theta) \Bigr),
        \end{split}
    \end{equation}
    where \( \hat{\theta} \) denotes the frozen parameters of \( \func \) from previous training iterations.}

    \item [(b)] {For images with high idempotency but low quality, we introduce a compactness objective to prevent \( \func \) from mapping to these undesirable results.  
    Specifically, given a fixed initial mapping \( \func(\comp; \hat{\theta}) \), we maximize its distance from its second mapping to reduce its idempotency score:
    \begin{equation}
        \begin{split}
            \theta^{*} &= \argmax_{\theta} \losscomp \Bigl( \func \bigl( \func(\comp; \hat{\theta}); \theta \bigr), \func(\comp; \hat{\theta}) \Bigr)\\
            &= \argmax_{\theta} \mathcal{D} \Bigl( \func \bigl( \func(\comp; \hat{\theta}); \theta \bigr), \func(\comp; \hat{\theta}) \Bigr).
        \end{split}
    \end{equation}}

    \item [(c)] {In practice, the compactness objective can be unstable due to its maximization nature and large gradients.  
    To improve training stability, we constrain the compactness objective within a dynamic bound tailored to each image.  
    The improved compactness objective \( \losscompp \) is defined as:
    \begin{equation}
        \losscompp = \tanh \left( \frac{\losscomp}{a \lossiden} \right) \cdot a \lossiden,
    \end{equation}
    where \( a \lossiden \) serves as the upper bound for the compactness term.  
    When \( \losscomp \ll a \lossiden \), the tanh function behaves linearly and \( \losscompp \approx \losscomp \).  
    When \( \losscomp \gg a \lossiden \), the function saturates and \( \losscompp \approx a \lossiden \).  
    This formulation ensures smoother optimization by reducing sensitivity to large gradients.}
\end{itemize}

\subsubsection{Overall objective}
Combining the four objectives described above, we define the overall loss function of our proposed domain-consistent quality enhancement method as:
\begin{equation}
    \begin{split}
        \loss (\theta, \hat{\theta}) &= \lossenh (\theta) + \widen \lossiden (\theta)\\
        &+ \widem \lossidem (\theta, \hat{\theta}) - \wcomp \losscompp (\theta, \hat{\theta}).
    \end{split}
\end{equation}
Here, \( \widen \), \( \widem \), and \( \wcomp \) are the weights for the identity, idempotency, and compactness objectives, respectively.
The complete training procedure is summarized in \cref{alg:training}.

\begin{algorithm}[!t]
\caption{Training Routine for Domain-Consistent Quality Enhancement}
\vspace{3pt}\textbf{StopGrad}($\cdot$) prevents gradient back-propagation through its input argument.
\label{alg:training}
\begin{algorithmic}[1]
    \Require Image batch \( \{ (\comp^i, \raw^i) \}_{i=1}^{B} \), hyper-parameters \( \widen \), \( \widem \), \( \wcomp \), \( a \), learning rate \( \eta \)
    \State Initialize model parameters \( \theta \)
    \For{each training iteration}
        \Statex \textbf{Forward pass:}
        \State \( \enh^i \gets \func (\comp^i, \theta) \) for \( i = 1, \ldots, B \)
        \State \( \rawenh^i \gets \func (\raw^i, \theta) \) for \( i = 1, \ldots, B \)
        \State \( \enhenh^i \gets \func (\enh^i, \theta) \) for \( i = 1, \ldots, B \)
        \State \( \enhdenh^i \gets \func \bigl( \text{StopGrad}(\enh^i), \theta \bigr) \) for \( i = 1, \ldots, B \)

        \Statex \textbf{Loss computation:}
        \State \( \lossenh \gets \mathbb{E}_i \left[ \mathcal{D}(\enh^i, \raw^i) \right] \)
        \State \( \lossiden \gets \mathbb{E}_i \left[ \mathcal{D}(\rawenh^i, \raw^i) \right] \)
        \State \( \lossidem \gets \mathbb{E}_i \left[ \mathcal{D} \left( \text{StopGrad}(\enhenh^i), \enh^i \right) \right] \)
        \State \( \losscomp \gets \mathbb{E}_i \left[ \mathcal{D} \left( \enhdenh^i, \text{StopGrad}(\enh^i) \right) \right] \)
        \State \( \losscompp \gets \tanh \left( \frac{\losscomp}{a \lossiden} \right) \cdot a \lossiden \)
        \State \( \loss \gets \lossenh + \widen \lossiden + \widem \lossidem - \wcomp \losscompp \)

        \Statex \textbf{Parameter update:}
        \State \( \theta \gets \theta - \eta \nabla_\theta \loss \)
    \EndFor
    \Ensure Learned parameters \( \theta \)
\end{algorithmic}
\end{algorithm}

\subsection{Theoretical Analysis}

In this section, we provide a theoretical analysis of our proposed domain-consistent quality enhancement method, focusing on the four objectives discussed above.

\begin{theorem}[Alignment with Natural Domain]
    We define the generated domain, denoted by \( \denh \), as the Probability Density Function (PDF) of the enhanced images \( \enh = \func ( \comp; \theta) \), where \( \comp \sim \dcomp \).
    We decompose the identity, idempotency, and compactness objectives into two terms:
    \begin{equation}
        \loss' (\theta, \hat{\theta}) = \underbrace{\lossiden (\theta) - \lambda_c \losscompp (\theta, \hat{\theta})}_{\lossgroup} + \lambda_i \lossidem (\theta, \hat{\theta}),
    \end{equation}
    where \( \lambda_c = \wcomp / \widen \) and \( \lambda_i = \widem / \widen \).
    Assuming a sufficiently large model capacity, both terms can attain a global minimum:
    \begin{equation}
        \label{eq:assume}
        \theta^* = \argmin_{\theta} \lossgroup (\theta, {\theta^*}) = \argmin_{\theta} \lossidem (\theta, {\theta^*}).
    \end{equation}
    Then, there exists a \( \theta^* \) such that \( \denhbest = \dnat \), and for \( \lambda_c = 1 \), this is the only possible \( \denhbest \).
\end{theorem}

We prove the above theorem in Appendix~\ref{app:1}.
Theorem~1 shows that the identity, idempotency, and compactness objectives work together to ensure that the generated domain aligns with the natural domain, even in the absence of paired samples \( (\comp, \raw) \).
In addition, the enhancement objective ensures that the enhanced image \( \enh \) is faithful to the corresponding raw image \( \raw \) when pairs are available.
Together, these objectives collectively enable effective domain-consistent quality enhancement.

In practice, we use \( \lambda_c < 1 \).
Although the theoretical analysis guarantees a unique desired optimum when \( \lambda_c = 1 \), a neural network with finite capacity may suffer from training instability, as optimizing \( \losscomp \) can lead to large gradients---even for its improved formulation.

\begin{table*}[!t]
    \caption{Degradation Index (DI) Values and Relative Reduction (\%) of Adapted Methods Compared to Baselines on the DIV2K Dataset}
    \label{tab:di-improvement}
    
    \centering
    \begin{tabular}{c @{} c >{\hspace{-7pt}} c >{\hspace{-7pt}} c >{\hspace{-7pt}} c >{\hspace{-7pt}} c >{\hspace{-7pt}} c >{\hspace{-7pt}} c >{\hspace{-7pt}} c >{\hspace{-7pt}} c >{\hspace{-7pt}} c}
        \toprule
        QP/QF & \cite{dongCompressionArtifactsReduction2015} & \cite{zhangGaussianDenoiserResidual2017} & \cite{wangNovelDeepLearningbased2017} & \cite{zhangResidualDenseNetwork2018} & \cite{guoConvolutionalBlindDenoising2019} & \cite{xingEarlyExitNot2020a} & \cite{zamirMultistageProgressiveImage2021} & \cite{ledigPhotoRealisticSingleImage2017} & \cite{wangESRGANEnhancedSuperResolution2018} & \cite{wangRealESRGANTrainingRealWorld2021}\\
        \midrule
        \multicolumn{11}{c}{Codec: BPG / Metric: PSNR}\\
        \midrule
        27 & 1.97 (\red{-7\%}) & 1.13 (\red{-71\%}) & 1.20 (\red{-68\%}) & 2.52 (\red{-28\%}) & 0.86 (\red{-79\%}) & 0.22 (\red{-95\%}) & 0.64 (\red{-86\%}) & 7.01 (\red{-17\%}) & 7.13 (\red{-16\%}) & 7.24 (\red{-6\%})\\
        32 & 1.68 (\red{-5\%}) & 0.92 (\red{-66\%}) & 1.44 (\red{-39\%}) & 2.11 (\red{-14\%}) & 0.97 (\red{-72\%}) & 0.44 (\red{-88\%}) & 0.93 (\red{-71\%}) & 4.25 (\red{-52\%}) & 4.29 (\red{-56\%}) & 6.18 (\red{-5\%})\\
        37 & 1.62 (\red{-5\%}) & 0.29 (\red{-86\%}) & 1.68 (\red{-19\%}) & 2.06 (\red{-7\%}) & 1.13 (\red{-64\%}) & 0.43 (\red{-84\%}) & 1.57 (\red{-40\%}) & 3.19 (\red{-64\%}) & 3.85 (\red{-61\%}) & 2.03 (\red{-68\%})\\
        42 & 1.71 (\red{-6\%}) & 1.40 (\red{-31\%}) & 1.74 (\red{-15\%}) & 1.99 (\red{-7\%}) & 1.73 (\red{-45\%}) & 0.32 (\red{-87\%}) & 2.25 (\red{-13\%}) & 2.75 (\red{-64\%}) & 3.44 (\red{-61\%}) & 0.42 (\red{-94\%})\\
        Ave. & 1.75 (\red{-6\%}) & 0.93 (\red{-65\%}) & 1.52 (\red{-40\%}) & 2.17 (\red{-16\%}) & 1.17 (\red{-66\%}) & 0.35 (\red{-89\%}) & 1.35 (\red{-59\%}) & 4.30 (\red{-49\%}) & 4.68 (\red{-49\%}) & 3.97 (\red{-42\%})\\
        \midrule
        \multicolumn{11}{c}{Codec: BPG / Metric: SSIM}\\
        \midrule
        27 & 0.15 (\red{-10\%}) & 0.15 (\red{-45\%}) & 0.17 (\red{-35\%}) & 0.21 (\red{-14\%}) & 0.07 (\red{-79\%}) & 0.02 (\red{-95\%}) & 0.03 (\red{-91\%}) & 7.45 (\red{-36\%}) & 7.39 (\red{-34\%}) & 7.25 (\red{-19\%})\\
        32 & 0.27 (\red{-9\%}) & 0.26 (\red{-49\%}) & 0.35 (\red{-25\%}) & 0.42 (\red{-7\%}) & 0.17 (\red{-74\%}) & 0.10 (\red{-82\%}) & 0.12 (\red{-80\%}) & 3.40 (\red{-72\%}) & 3.32 (\red{-76\%}) & 5.21 (\red{-25\%})\\
        37 & 0.52 (\red{-7\%}) & 0.14 (\red{-84\%}) & 0.68 (\red{-13\%}) & 0.78 (\red{-5\%}) & 0.36 (\red{-66\%}) & 0.14 (\red{-85\%}) & 0.48 (\red{-51\%}) & 2.30 (\red{-82\%}) & 3.31 (\red{-78\%}) & 3.28 (\red{-53\%})\\
        42 & 0.90 (\red{-6\%}) & 0.86 (\red{-28\%}) & 1.01 (\red{-12\%}) & 1.19 (\red{-4\%}) & 0.91 (\red{-45\%}) & 0.14 (\red{-90\%}) & 1.22 (\red{-20\%}) & 2.54 (\red{-73\%}) & 2.57 (\red{-80\%}) & 0.41 (\red{-94\%})\\
        Ave. & 0.46 (\red{-7\%}) & 0.35 (\red{-50\%}) & 0.55 (\red{-17\%}) & 0.65 (\red{-6\%}) & 0.38 (\red{-59\%}) & 0.10 (\red{-88\%}) & 0.46 (\red{-46\%}) & 3.92 (\red{-66\%}) & 4.15 (\red{-69\%}) & 4.04 (\red{-47\%})\\
        \midrule
        \multicolumn{11}{c}{Codec: BPG / Metric: TOPIQ}\\
        \midrule
        27 & 1.25 (\red{-7\%}) & 1.17 (\red{-59\%}) & 1.21 (\red{-55\%}) & 2.05 (\red{-26\%}) & 0.69 (\red{-80\%}) & 0.10 (\red{-97\%}) & 0.39 (\red{-89\%}) & 9.15 (\red{-16\%}) & 9.61 (\red{-18\%}) & 8.35 (\red{-11\%})\\
        32 & 2.34 (\red{-5\%}) & 2.32 (\red{-45\%}) & 2.92 (\red{-22\%}) & 3.56 (\red{-8\%}) & 1.86 (\red{-65\%}) & 0.79 (\red{-84\%}) & 1.34 (\red{-75\%}) & 5.48 (\red{-53\%}) & 5.66 (\red{-57\%}) & 7.64 (\red{-8\%})\\
        37 & 3.32 (\red{-5\%}) & 1.27 (\red{-74\%}) & 4.26 (\red{-9\%}) & 4.74 (\red{-4\%}) & 2.72 (\red{-54\%}) & 0.84 (\red{-84\%}) & 3.44 (\red{-39\%}) & 3.71 (\red{-70\%}) & 5.57 (\red{-58\%}) & 4.45 (\red{-40\%})\\
        42 & 2.98 (\red{-5\%}) & 3.09 (\red{-20\%}) & 3.49 (\red{-8\%}) & 3.86 (\red{-3\%}) & 3.17 (\red{-34\%}) & 0.24 (\red{-94\%}) & 4.02 (\red{-14\%}) & 3.04 (\red{-74\%}) & 4.95 (\red{-61\%}) & 0.01 (\red{-100\%})\\
        Ave. & 2.47 (\red{-5\%}) & 1.96 (\red{-50\%}) & 2.97 (\red{-20\%}) & 3.55 (\red{-9\%}) & 2.11 (\red{-56\%}) & 0.49 (\red{-89\%}) & 2.30 (\red{-52\%}) & 5.34 (\red{-54\%}) & 6.45 (\red{-49\%}) & 5.11 (\red{-38\%})\\
        \midrule
        \multicolumn{11}{c}{Codec: BPG / Metric: LPIPS}\\
        \midrule
        27 & 8.25 (\red{-7\%}) & 5.42 (\red{-51\%}) & 5.58 (\red{-47\%}) & 9.05 (\red{-17\%}) & 2.69 (\red{-81\%}) & 0.45 (\red{-97\%}) & 1.39 (\red{-91\%}) & 119.44 (\red{-37\%}) & 120.07 (\red{-32\%}) & 149.89 (\red{-8\%})\\
        32 & 5.62 (\red{-7\%}) & 3.22 (\red{-57\%}) & 4.62 (\red{-30\%}) & 6.32 (\red{-10\%}) & 2.17 (\red{-77\%}) & 0.86 (\red{-91\%}) & 2.03 (\red{-81\%}) & 28.41 (\red{-79\%}) & 29.10 (\red{-80\%}) & 56.71 (\red{-21\%})\\
        37 & 3.68 (\red{-7\%}) & 0.36 (\red{-92\%}) & 3.83 (\red{-13\%}) & 4.61 (\red{-5\%}) & 1.89 (\red{-67\%}) & 0.44 (\red{-92\%}) & 3.05 (\red{-48\%}) & 11.62 (\red{-85\%}) & 17.44 (\red{-83\%}) & 15.21 (\red{-64\%})\\
        42 & 2.15 (\red{-8\%}) & 2.07 (\red{-28\%}) & 2.51 (\red{-10\%}) & 2.62 (\red{-2\%}) & 1.86 (\red{-41\%}) & 0.04 (\red{-99\%}) & 2.62 (\red{-13\%}) & 8.72 (\red{-78\%}) & 11.72 (\red{-79\%}) & -0.25 (\red{-101\%})\\
        Ave. & 4.93 (\red{-7\%}) & 2.77 (\red{-58\%}) & 4.14 (\red{-32\%}) & 5.65 (\red{-11\%}) & 2.15 (\red{-73\%}) & 0.45 (\red{-94\%}) & 2.27 (\red{-74\%}) & 42.05 (\red{-62\%}) & 44.58 (\red{-63\%}) & 55.39 (\red{-29\%})\\
        \midrule
        \multicolumn{11}{c}{Codec: JPEG / Metric: PSNR}\\
        \midrule
        30 & 2.19 (\red{-2\%}) & 1.57 (\red{-40\%}) & 1.66 (\red{-36\%}) & 2.69 (\red{-6\%}) & 0.09 (\red{-97\%}) & 0.31 (\red{-90\%}) & 2.80 (\red{-12\%}) & 4.09 (\red{-41\%}) & 4.83 (\red{-27\%}) & 2.12 (\red{-61\%})\\
        40 & 2.19 (\red{-2\%}) & 1.35 (\red{-50\%}) & 1.84 (\red{-33\%}) & 2.70 (\red{-8\%}) & 0.16 (\red{-95\%}) & 0.11 (\red{-96\%}) & 1.63 (\red{-52\%}) & 4.40 (\red{-35\%}) & 5.03 (\red{-29\%}) & 3.75 (\red{-36\%})\\
        50 & 2.33 (\red{-3\%}) & 1.30 (\red{-53\%}) & 1.62 (\red{-45\%}) & 2.75 (\red{-9\%}) & 0.14 (\red{-96\%}) & 0.07 (\red{-98\%}) & 0.64 (\red{-80\%}) & 4.43 (\red{-35\%}) & 4.96 (\red{-34\%}) & 4.65 (\red{-9\%})\\
        60 & 2.50 (\red{-3\%}) & 1.45 (\red{-55\%}) & 1.51 (\red{-56\%}) & 2.91 (\red{-11\%}) & 0.10 (\red{-97\%}) & 0.10 (\red{-98\%}) & 1.01 (\red{-73\%}) & 4.96 (\red{-31\%}) & 5.05 (\red{-27\%}) & 5.08 (\red{-11\%})\\
        Ave. & 2.30 (\red{-3\%}) & 1.42 (\red{-50\%}) & 1.66 (\red{-43\%}) & 2.76 (\red{-8\%}) & 0.12 (\red{-97\%}) & 0.15 (\red{-96\%}) & 1.52 (\red{-55\%}) & 4.47 (\red{-35\%}) & 4.97 (\red{-29\%}) & 3.90 (\red{-29\%})\\
        \midrule
        \multicolumn{11}{c}{Codec: JPEG / Metric: SSIM}\\
        \midrule
        30 & 1.25 (\red{-2\%}) & 1.01 (\red{-28\%}) & 1.07 (\red{-24\%}) & 1.48 (\red{-1\%}) & 0.02 (\red{-99\%}) & 0.11 (\red{-92\%}) & 1.57 (\red{-16\%}) & 2.53 (\red{-64\%}) & 3.41 (\red{-54\%}) & 2.62 (\red{-50\%})\\
        40 & 1.05 (\red{-2\%}) & 0.83 (\red{-33\%}) & 0.96 (\red{-20\%}) & 1.30 (0\%) & 0.02 (\red{-98\%}) & 0.02 (\red{-98\%}) & 0.65 (\red{-60\%}) & 3.01 (\red{-58\%}) & 4.35 (\red{-47\%}) & 4.94 (\red{-30\%})\\
        50 & 0.91 (\red{-2\%}) & 0.68 (\red{-36\%}) & 0.74 (\red{-29\%}) & 1.07 (\red{-1\%}) & 0.03 (\red{-98\%}) & 0.01 (\red{-99\%}) & 0.06 (\red{-95\%}) & 3.11 (\red{-53\%}) & 3.41 (\red{-63\%}) & -0.17 (\red{-105\%})\\
        60 & 0.75 (\red{-2\%}) & 0.64 (\red{-32\%}) & 0.61 (\red{-34\%}) & 0.96 (\red{-1\%}) & 0.02 (\red{-98\%}) & 0.01 (\red{-99\%}) & 0.14 (\red{-88\%}) & 3.54 (\red{-51\%}) & 3.78 (\red{-51\%}) & 0.20 (\red{-95\%})\\
        Ave. & 0.99 (\red{-2\%}) & 0.79 (\red{-32\%}) & 0.85 (\red{-26\%}) & 1.20 (\red{-1\%}) & 0.02 (\red{-98\%}) & 0.04 (\red{-97\%}) & 0.60 (\red{-59\%}) & 3.05 (\red{-56\%}) & 3.74 (\red{-54\%}) & 1.90 (\red{-63\%})\\
        \midrule
        \multicolumn{11}{c}{Codec: JPEG / Metric: TOPIQ}\\
        \midrule
        30 & 6.16 (\red{-1\%}) & 5.01 (\red{-21\%}) & 5.23 (\red{-19\%}) & 6.64 (\red{-2\%}) & 0.06 (\red{-99\%}) & 0.44 (\red{-94\%}) & 6.86 (\red{-11\%}) & 5.44 (\red{-49\%}) & 7.57 (\red{-35\%}) & 4.01 (\red{-57\%})\\
        40 & 4.80 (\red{-2\%}) & 3.79 (\red{-30\%}) & 4.31 (\red{-21\%}) & 5.60 (\red{-2\%}) & 0.06 (\red{-99\%}) & 0.09 (\red{-99\%}) & 3.63 (\red{-46\%}) & 5.21 (\red{-53\%}) & 7.29 (\red{-36\%}) & 6.47 (\red{-26\%})\\
        50 & 3.99 (\red{-2\%}) & 2.77 (\red{-38\%}) & 3.15 (\red{-33\%}) & 4.63 (\red{-4\%}) & 0.06 (\red{-99\%}) & 0.03 (\red{-99\%}) & 0.52 (\red{-90\%}) & 5.61 (\red{-44\%}) & 6.66 (\red{-39\%}) & 5.81 (\red{-21\%})\\
        60 & 3.29 (\red{-3\%}) & 2.26 (\red{-47\%}) & 2.32 (\red{-48\%}) & 3.89 (\red{-9\%}) & 0.04 (\red{-99\%}) & 0.03 (\red{-100\%}) & 0.96 (\red{-82\%}) & 6.03 (\red{-35\%}) & 5.94 (\red{-41\%}) & 5.65 (\red{-26\%})\\
        Ave. & 4.56 (\red{-2\%}) & 3.46 (\red{-32\%}) & 3.75 (\red{-29\%}) & 5.19 (\red{-4\%}) & 0.05 (\red{-99\%}) & 0.15 (\red{-98\%}) & 2.99 (\red{-52\%}) & 5.57 (\red{-46\%}) & 6.86 (\red{-38\%}) & 5.49 (\red{-33\%})\\
        \midrule
        \multicolumn{11}{c}{Codec: JPEG / Metric: LPIPS}\\
        \midrule
        30 & 14.45 (\red{-2\%}) & 11.77 (\red{-29\%}) & 11.92 (\red{-26\%}) & 16.39 (\red{-2\%}) & 0.14 (\red{-99\%}) & 0.68 (\red{-96\%}) & 19.70 (\red{-7\%}) & 29.85 (\red{-66\%}) & 42.30 (\red{-55\%}) & 21.45 (\red{-73\%})\\
        40 & 16.03 (\red{-2\%}) & 12.53 (\red{-35\%}) & 14.79 (\red{-22\%}) & 19.40 (\red{-1\%}) & 0.16 (\red{-99\%}) & 0.30 (\red{-99\%}) & 9.98 (\red{-61\%}) & 49.29 (\red{-59\%}) & 66.17 (\red{-48\%}) & 67.68 (\red{-49\%})\\
        50 & 18.91 (\red{-2\%}) & 14.37 (\red{-35\%}) & 15.42 (\red{-30\%}) & 21.80 (\red{-1\%}) & 0.77 (\red{-97\%}) & 0.21 (\red{-99\%}) & 1.64 (\red{-94\%}) & 56.69 (\red{-61\%}) & 56.40 (\red{-66\%}) & 60.36 (\red{-40\%})\\
        60 & 21.69 (\red{-2\%}) & 17.97 (\red{-33\%}) & 17.61 (\red{-34\%}) & 25.64 (\red{-3\%}) & 0.98 (\red{-97\%}) & 0.28 (\red{-99\%}) & 6.39 (\red{-80\%}) & 72.69 (\red{-56\%}) & 77.54 (\red{-56\%}) & 77.25 (\red{-45\%})\\
        Ave. & 17.77 (\red{-2\%}) & 14.16 (\red{-33\%}) & 14.93 (\red{-29\%}) & 20.81 (\red{-1\%}) & 0.51 (\red{-98\%}) & 0.37 (\red{-98\%}) & 9.43 (\red{-64\%}) & 52.13 (\red{-60\%}) & 60.60 (\red{-57\%}) & 56.69 (\red{-50\%})\\
        \bottomrule
    \end{tabular}
\end{table*}

\section{Experiments}

In this section, we conduct extensive experiments to evaluate the performance of our proposed domain-consistent quality enhancement method.
The goal of our experiments is to demonstrate the effectiveness of our proposed method in addressing the multi-enhancement challenges.

\subsection{Experimental Setup}

\subsubsection*{\textbf{Datasets}}

Following the setup in \cref{sec:findings}, we conduct our experiments on the DIV2K and Flickr2K datasets.
All images are compressed using two widely adopted codecs: the commonly used JPEG codec and the more efficient BPG codec.
Note that BPG implements the HEVC-MSP standard.
Each codec is applied at four compression levels: JPEG uses Quality Factors (QF) of 30, 40, 50, and 60, while BPG uses Quantization Parameters (QP) of 27, 32, 37, and 42.

\subsubsection*{\textbf{Baselines}}

Compressed images are enhanced using several prevalent quality enhancement baselines, including the previously mentioned AR-CNN, DnCNN, DCAD, and RBQE.
We also evaluate more advanced image enhancement methods, such as RDN, CBDNet, and MPRNet.
Additionally, we include GAN-based methods, such as SRGAN, ESRGAN, and Real-ESRGAN, which are widely used for perceptual quality enhancement.
All baseline methods are trained on the DIV2K training set and evaluated on the DIV2K validation set as well as the last 100 images of the Flickr2K dataset.

\begin{table*}[!t]
    \caption{Degradation Index (DI) Values and Relative Reduction (\%) of Adapted Methods Compared to Baselines on the Flickr2K Dataset}
    \label{tab:di-improvement-flickr2k}
    
    \centering
    \begin{tabular}{c @{} c >{\hspace{-7.5pt}} c >{\hspace{-7.5pt}} c >{\hspace{-7.5pt}} c >{\hspace{-7.5pt}} c >{\hspace{-7.5pt}} c >{\hspace{-7.5pt}} c >{\hspace{-7.5pt}} c >{\hspace{-7.5pt}} c >{\hspace{-7.5pt}} c}
        \toprule
        QP/QF & \cite{dongCompressionArtifactsReduction2015} & \cite{zhangGaussianDenoiserResidual2017} & \cite{wangNovelDeepLearningbased2017} & \cite{zhangResidualDenseNetwork2018} & \cite{guoConvolutionalBlindDenoising2019} & \cite{xingEarlyExitNot2020a} & \cite{zamirMultistageProgressiveImage2021} & \cite{ledigPhotoRealisticSingleImage2017} & \cite{wangESRGANEnhancedSuperResolution2018} & \cite{wangRealESRGANTrainingRealWorld2021}\\
        \midrule
        \multicolumn{11}{c}{Codec: BPG / Metric: PSNR}\\
        \midrule
        27 & 2.17 (\red{-6\%}) & 1.28 (\red{-70\%}) & 1.38 (\red{-67\%}) & 2.89 (\red{-26\%}) & 0.93 (\red{-80\%}) & 0.26 (\red{-95\%}) & 0.69 (\red{-87\%}) & 7.44 (\red{-15\%}) & 7.40 (\red{-15\%}) & 7.84 (\red{-6\%})\\
        32 & 1.89 (\red{-5\%}) & 1.00 (\red{-67\%}) & 1.64 (\red{-37\%}) & 2.38 (\red{-13\%}) & 0.96 (\red{-76\%}) & 0.50 (\red{-88\%}) & 0.98 (\red{-74\%}) & 4.82 (\red{-47\%}) & 4.68 (\red{-53\%}) & 6.82 (\red{-2\%})\\
        37 & 1.85 (\red{-5\%}) & 0.33 (\red{-86\%}) & 1.86 (\red{-19\%}) & 2.29 (\red{-7\%}) & 1.12 (\red{-68\%}) & 0.46 (\red{-85\%}) & 1.53 (\red{-48\%}) & 3.59 (\red{-60\%}) & 4.29 (\red{-58\%}) & 2.49 (\red{-62\%})\\
        42 & 1.90 (\red{-6\%}) & 1.51 (\red{-33\%}) & 1.92 (\red{-14\%}) & 2.22 (\red{-7\%}) & 1.64 (\red{-52\%}) & 0.38 (\red{-86\%}) & 2.29 (\red{-20\%}) & 3.13 (\red{-62\%}) & 3.94 (\red{-57\%}) & 0.47 (\red{-93\%})\\
        Ave. & 1.95 (\red{-5\%}) & 1.03 (\red{-65\%}) & 1.70 (\red{-40\%}) & 2.45 (\red{-15\%}) & 1.16 (\red{-70\%}) & 0.40 (\red{-89\%}) & 1.37 (\red{-63\%}) & 4.75 (\red{-46\%}) & 5.08 (\red{-46\%}) & 4.41 (\red{-39\%})\\
        \midrule
        \multicolumn{11}{c}{Codec: BPG / Metric: SSIM}\\
        \midrule
        27 & 0.13 (\red{-9\%}) & 0.13 (\red{-47\%}) & 0.15 (\red{-35\%}) & 0.20 (\red{-12\%}) & 0.07 (\red{-77\%}) & 0.02 (\red{-94\%}) & 0.03 (\red{-88\%}) & 7.74 (\red{-32\%}) & 7.45 (\red{-32\%}) & 7.82 (\red{-17\%})\\
        32 & 0.28 (\red{-8\%}) & 0.24 (\red{-50\%}) & 0.35 (\red{-24\%}) & 0.42 (\red{-6\%}) & 0.16 (\red{-73\%}) & 0.09 (\red{-83\%}) & 0.10 (\red{-82\%}) & 3.85 (\red{-68\%}) & 3.65 (\red{-73\%}) & 5.18 (\red{-28\%})\\
        37 & 0.57 (\red{-7\%}) & 0.15 (\red{-84\%}) & 0.71 (\red{-14\%}) & 0.80 (\red{-5\%}) & 0.33 (\red{-69\%}) & 0.12 (\red{-87\%}) & 0.41 (\red{-57\%}) & 2.57 (\red{-80\%}) & 3.56 (\red{-76\%}) & 3.70 (\red{-47\%})\\
        42 & 1.02 (\red{-6\%}) & 0.95 (\red{-29\%}) & 1.16 (\red{-12\%}) & 1.29 (\red{-4\%}) & 0.83 (\red{-53\%}) & 0.14 (\red{-90\%}) & 1.23 (\red{-27\%}) & 2.74 (\red{-73\%}) & 2.94 (\red{-78\%}) & 0.48 (\red{-94\%})\\
        Ave. & 0.50 (\red{-7\%}) & 0.37 (\red{-50\%}) & 0.59 (\red{-16\%}) & 0.68 (\red{-5\%}) & 0.35 (\red{-63\%}) & 0.09 (\red{-88\%}) & 0.44 (\red{-49\%}) & 4.22 (\red{-64\%}) & 4.40 (\red{-67\%}) & 4.30 (\red{-46\%})\\
        \midrule
        \multicolumn{11}{c}{Codec: BPG / Metric: TOPIQ}\\
        \midrule
        27 & 1.29 (\red{-7\%}) & 1.19 (\red{-61\%}) & 1.21 (\red{-57\%}) & 2.20 (\red{-26\%}) & 0.75 (\red{-79\%}) & 0.13 (\red{-97\%}) & 0.48 (\red{-88\%}) & 9.65 (\red{-13\%}) & 9.97 (\red{-15\%}) & 8.78 (\red{-11\%})\\
        32 & 2.48 (\red{-5\%}) & 2.36 (\red{-48\%}) & 2.99 (\red{-24\%}) & 3.78 (\red{-9\%}) & 1.91 (\red{-66\%}) & 0.86 (\red{-85\%}) & 1.39 (\red{-75\%}) & 6.09 (\red{-49\%}) & 5.83 (\red{-55\%}) & 7.85 (\red{-10\%})\\
        37 & 3.79 (\red{-4\%}) & 1.43 (\red{-74\%}) & 4.69 (\red{-10\%}) & 5.29 (\red{-4\%}) & 2.79 (\red{-56\%}) & 0.95 (\red{-84\%}) & 3.32 (\red{-47\%}) & 4.27 (\red{-66\%}) & 6.09 (\red{-55\%}) & 5.14 (\red{-33\%})\\
        42 & 3.73 (\red{-4\%}) & 3.70 (\red{-20\%}) & 4.27 (\red{-8\%}) & 4.70 (\red{-3\%}) & 3.41 (\red{-40\%}) & 0.41 (\red{-92\%}) & 4.46 (\red{-20\%}) & 3.41 (\red{-72\%}) & 5.43 (\red{-58\%}) & -0.11 (\red{-101\%})\\
        Ave. & 2.82 (\red{-5\%}) & 2.17 (\red{-51\%}) & 3.29 (\red{-21\%}) & 3.99 (\red{-8\%}) & 2.22 (\red{-58\%}) & 0.59 (\red{-89\%}) & 2.41 (\red{-55\%}) & 5.86 (\red{-51\%}) & 6.83 (\red{-47\%}) & 5.42 (\red{-37\%})\\
        \midrule
        \multicolumn{11}{c}{Codec: BPG / Metric: LPIPS}\\
        \midrule
        27 & 9.87 (\red{-6\%}) & 6.67 (\red{-53\%}) & 6.89 (\red{-48\%}) & 11.74 (\red{-16\%}) & 3.26 (\red{-82\%}) & 0.62 (\red{-96\%}) & 2.15 (\red{-89\%}) & 138.75 (\red{-33\%}) & 130.51 (\red{-32\%}) & 170.04 (\red{-7\%})\\
        32 & 6.93 (\red{-6\%}) & 3.80 (\red{-61\%}) & 5.82 (\red{-29\%}) & 8.00 (\red{-10\%}) & 2.62 (\red{-80\%}) & 1.15 (\red{-91\%}) & 2.61 (\red{-82\%}) & 35.67 (\red{-75\%}) & 34.17 (\red{-76\%}) & 69.60 (\red{-17\%})\\
        37 & 4.86 (\red{-7\%}) & 0.62 (\red{-90\%}) & 4.93 (\red{-15\%}) & 6.06 (\red{-5\%}) & 2.30 (\red{-73\%}) & 0.65 (\red{-92\%}) & 3.59 (\red{-58\%}) & 16.14 (\red{-81\%}) & 21.77 (\red{-80\%}) & 19.14 (\red{-60\%})\\
        42 & 3.13 (\red{-8\%}) & 2.64 (\red{-30\%}) & 3.38 (\red{-11\%}) & 3.70 (\red{-3\%}) & 2.20 (\red{-52\%}) & 0.21 (\red{-95\%}) & 3.42 (\red{-20\%}) & 11.32 (\red{-76\%}) & 14.55 (\red{-76\%}) & -0.20 (\red{-101\%})\\
        Ave. & 6.20 (\red{-6\%}) & 3.43 (\red{-59\%}) & 5.25 (\red{-32\%}) & 7.37 (\red{-11\%}) & 2.59 (\red{-77\%}) & 0.66 (\red{-94\%}) & 2.94 (\red{-75\%}) & 50.47 (\red{-58\%}) & 50.25 (\red{-60\%}) & 64.65 (\red{-26\%})\\
        \midrule
        \multicolumn{11}{c}{Codec: JPEG / Metric: PSNR}\\
        \midrule
        30 & 2.37 (\red{-2\%}) & 1.72 (\red{-39\%}) & 1.83 (\red{-35\%}) & 2.90 (\red{-5\%}) & 0.06 (\red{-98\%}) & 0.37 (\red{-89\%}) & 2.97 (\red{-15\%}) & 4.50 (\red{-39\%}) & 5.37 (\red{-27\%}) & 2.40 (\red{-59\%})\\
        40 & 2.37 (\red{-2\%}) & 1.48 (\red{-49\%}) & 2.04 (\red{-32\%}) & 2.88 (\red{-7\%}) & 0.08 (\red{-98\%}) & 0.13 (\red{-96\%}) & 1.84 (\red{-53\%}) & 4.84 (\red{-34\%}) & 5.54 (\red{-27\%}) & 4.14 (\red{-35\%})\\
        50 & 2.52 (\red{-2\%}) & 1.39 (\red{-53\%}) & 1.76 (\red{-44\%}) & 2.98 (\red{-9\%}) & 0.05 (\red{-99\%}) & 0.10 (\red{-97\%}) & 0.63 (\red{-83\%}) & 5.03 (\red{-30\%}) & 5.40 (\red{-33\%}) & 4.87 (\red{-11\%})\\
        60 & 2.69 (\red{-3\%}) & 1.63 (\red{-54\%}) & 1.72 (\red{-54\%}) & 3.19 (\red{-10\%}) & 0.06 (\red{-99\%}) & 0.10 (\red{-98\%}) & 1.06 (\red{-76\%}) & 5.49 (\red{-27\%}) & 5.48 (\red{-25\%}) & 5.41 (\red{-12\%})\\
        Ave. & 2.49 (\red{-2\%}) & 1.55 (\red{-49\%}) & 1.84 (\red{-42\%}) & 2.99 (\red{-8\%}) & 0.06 (\red{-98\%}) & 0.18 (\red{-95\%}) & 1.63 (\red{-58\%}) & 4.96 (\red{-33\%}) & 5.45 (\red{-28\%}) & 4.20 (\red{-29\%})\\
        \midrule
        \multicolumn{11}{c}{Codec: JPEG / Metric: SSIM}\\
        \midrule
        30 & 1.36 (\red{-2\%}) & 1.11 (\red{-29\%}) & 1.19 (\red{-25\%}) & 1.59 (\red{-1\%}) & 0.02 (\red{-99\%}) & 0.11 (\red{-93\%}) & 1.62 (\red{-20\%}) & 2.77 (\red{-58\%}) & 3.86 (\red{-49\%}) & 2.88 (\red{-49\%})\\
        40 & 1.11 (\red{-2\%}) & 0.88 (\red{-35\%}) & 1.04 (\red{-21\%}) & 1.36 (\red{-1\%}) & 0.02 (\red{-99\%}) & 0.03 (\red{-98\%}) & 0.68 (\red{-60\%}) & 3.27 (\red{-54\%}) & 4.64 (\red{-43\%}) & 5.32 (\red{-25\%})\\
        50 & 0.97 (\red{-2\%}) & 0.71 (\red{-37\%}) & 0.77 (\red{-31\%}) & 1.12 (\red{-1\%}) & 0.01 (\red{-99\%}) & 0.01 (\red{-99\%}) & 0.06 (\red{-95\%}) & 3.56 (\red{-45\%}) & 3.85 (\red{-58\%}) & -0.88 (\red{-125\%})\\
        60 & 0.79 (\red{-2\%}) & 0.67 (\red{-32\%}) & 0.64 (\red{-34\%}) & 0.97 (0\%) & 0.02 (\red{-98\%}) & 0.01 (\red{-99\%}) & 0.13 (\red{-89\%}) & 3.86 (\red{-45\%}) & 4.12 (\red{-47\%}) & -0.24 (\red{-106\%})\\
        Ave. & 1.06 (\red{-2\%}) & 0.84 (\red{-33\%}) & 0.91 (\red{-27\%}) & 1.26 (\red{-1\%}) & 0.02 (\red{-99\%}) & 0.04 (\red{-97\%}) & 0.62 (\red{-60\%}) & 3.37 (\red{-51\%}) & 4.12 (\red{-49\%}) & 1.77 (\red{-66\%})\\
        \midrule
        \multicolumn{11}{c}{Codec: JPEG / Metric: TOPIQ}\\
        \midrule
        30 & 6.64 (\red{-1\%}) & 5.29 (\red{-23\%}) & 5.55 (\red{-20\%}) & 7.10 (\red{-2\%}) & 0.06 (\red{-99\%}) & 0.59 (\red{-92\%}) & 6.81 (\red{-15\%}) & 5.94 (\red{-48\%}) & 8.37 (\red{-32\%}) & 4.46 (\red{-54\%})\\
        40 & 5.20 (\red{-1\%}) & 3.89 (\red{-33\%}) & 4.57 (\red{-22\%}) & 5.92 (\red{-3\%}) & 0.04 (\red{-99\%}) & 0.16 (\red{-98\%}) & 3.84 (\red{-46\%}) & 5.55 (\red{-53\%}) & 7.71 (\red{-35\%}) & 6.95 (\red{-25\%})\\
        50 & 4.35 (\red{-2\%}) & 2.78 (\red{-41\%}) & 3.18 (\red{-35\%}) & 4.92 (\red{-5\%}) & 0.04 (\red{-99\%}) & 0.07 (\red{-99\%}) & 0.55 (\red{-91\%}) & 6.12 (\red{-42\%}) & 7.07 (\red{-39\%}) & 6.49 (\red{-18\%})\\
        60 & 3.52 (\red{-3\%}) & 2.27 (\red{-50\%}) & 2.37 (\red{-51\%}) & 4.26 (\red{-9\%}) & 0.04 (\red{-99\%}) & 0.05 (\red{-99\%}) & 0.95 (\red{-84\%}) & 6.55 (\red{-34\%}) & 6.26 (\red{-41\%}) & 6.09 (\red{-25\%})\\
        Ave. & 4.93 (\red{-2\%}) & 3.56 (\red{-35\%}) & 3.92 (\red{-31\%}) & 5.55 (\red{-4\%}) & 0.05 (\red{-99\%}) & 0.22 (\red{-97\%}) & 3.04 (\red{-55\%}) & 6.04 (\red{-45\%}) & 7.35 (\red{-36\%}) & 6.00 (\red{-32\%})\\
        \midrule
        \multicolumn{11}{c}{Codec: JPEG / Metric: LPIPS}\\
        \midrule
        30 & 17.29 (\red{-2\%}) & 13.63 (\red{-31\%}) & 14.21 (\red{-28\%}) & 19.73 (\red{-2\%}) & 0.20 (\red{-99\%}) & 0.98 (\red{-96\%}) & 23.25 (\red{-13\%}) & 32.61 (\red{-64\%}) & 47.53 (\red{-51\%}) & 22.50 (\red{-74\%})\\
        40 & 19.20 (\red{-2\%}) & 14.20 (\red{-38\%}) & 17.42 (\red{-23\%}) & 22.72 (\red{-2\%}) & -0.00 (\red{-100\%}) & 0.47 (\red{-98\%}) & 12.97 (\red{-60\%}) & 54.32 (\red{-54\%}) & 70.67 (\red{-43\%}) & 66.29 (\red{-49\%})\\
        50 & 22.39 (\red{-2\%}) & 15.82 (\red{-38\%}) & 17.29 (\red{-33\%}) & 25.36 (\red{-2\%}) & 0.36 (\red{-99\%}) & 0.35 (\red{-99\%}) & 2.22 (\red{-93\%}) & 61.15 (\red{-56\%}) & 59.35 (\red{-62\%}) & 62.16 (\red{-35\%})\\
        60 & 25.36 (\red{-2\%}) & 19.82 (\red{-36\%}) & 19.55 (\red{-37\%}) & 29.71 (\red{-4\%}) & 0.65 (\red{-98\%}) & 0.36 (\red{-99\%}) & 6.26 (\red{-84\%}) & 73.04 (\red{-55\%}) & 78.38 (\red{-53\%}) & 84.60 (\red{-38\%})\\
        Ave. & 21.06 (\red{-2\%}) & 15.86 (\red{-36\%}) & 17.12 (\red{-31\%}) & 24.38 (\red{-3\%}) & 0.30 (\red{-99\%}) & 0.54 (\red{-98\%}) & 11.17 (\red{-65\%}) & 55.28 (\red{-57\%}) & 63.98 (\red{-53\%}) & 58.89 (\red{-48\%})\\
        \bottomrule
    \end{tabular}
\end{table*}

\begin{figure*}[!t]
    \centering
    \includegraphics[trim={0em 0em 0em 0em}, clip, width=\linewidth]{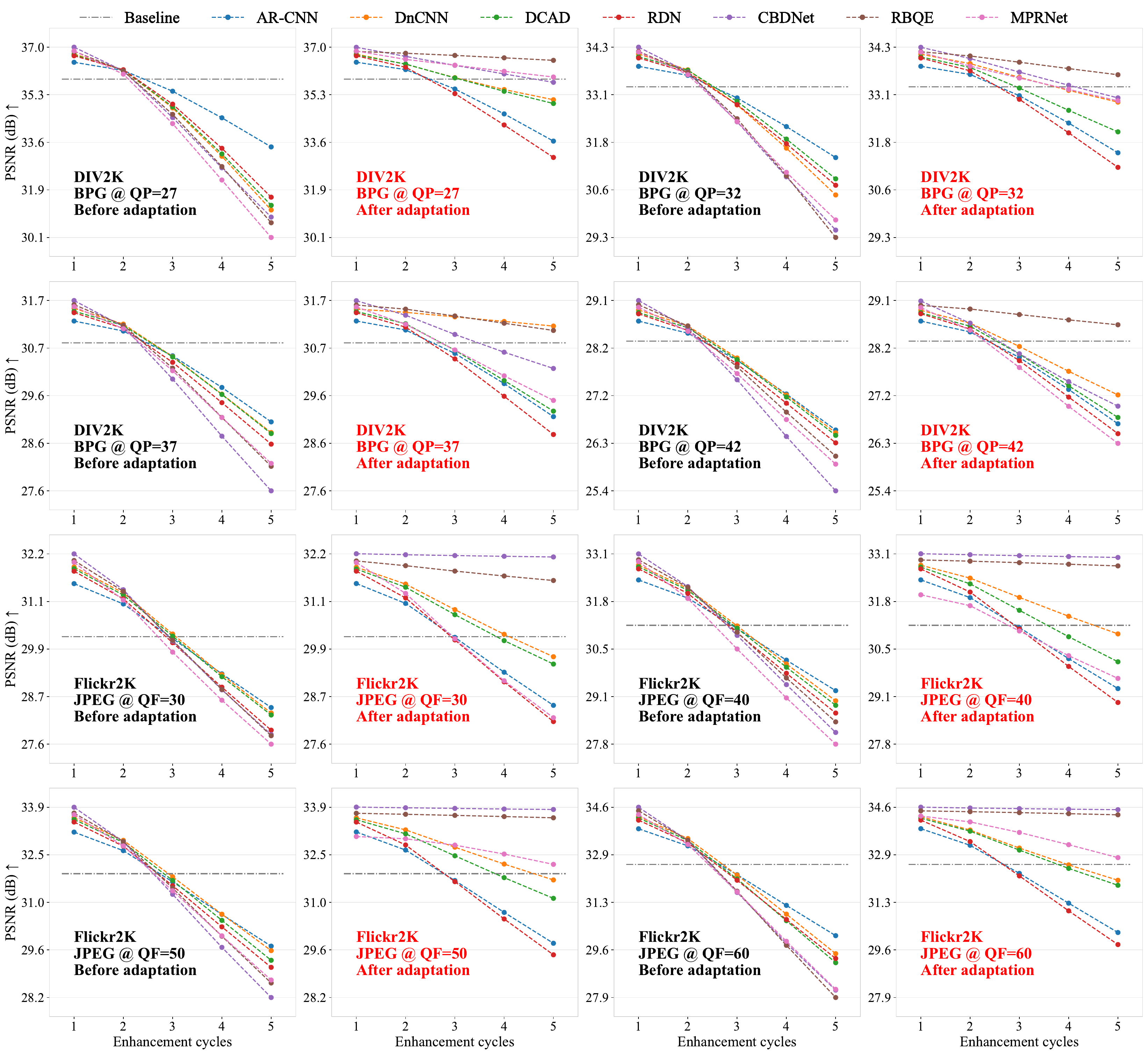}
    \caption{Objective results showing PSNR-based quality degradation under multi-enhancement across various datasets and codecs, evaluated using CNN-based methods before and after adaptation.} 
    \label{fig:quality-degradation-exp}
\end{figure*}

\subsubsection*{\textbf{Idempotent training}}

We adapt the baseline methods into domain-consistent quality enhancement ones by incorporating our proposed idempotent training objectives.
The training settings are kept consistent with the original methods.
The scaling factor \( a \) in the improved compactness objective is set to 1.5.
The distance metric \( \mathcal{D} \) and the weights for the objectives are configured as follows:
(1) For CNN-based methods, \( \mathcal{D} \) is the L1 loss function for all training objectives.
The weights for the identity (\( \widen \)), idempotency (\( \widem \)), and compactness (\( \wcomp \)) objectives are set to \( 1 \times 10^{-2} \), \( 1 \times 10^{-2} \), and \( 1 \times 10^{-3} \), respectively.
(2) For GAN-based methods, their enhancement objective typically includes three components: a pixel loss, a perceptual loss, and an adversarial loss~\cite{ledigPhotoRealisticSingleImage2017,wangESRGANEnhancedSuperResolution2018,wangRealESRGANTrainingRealWorld2021}.
For all objectives except the enhancement objective, \( \mathcal{D} \) is set to the perceptual loss. The weights \( \widen \), \( \widem \), and \( \wcomp \) are set to \( 1 \times 10^{-1} \), \( 1 \times 10^{-1} \), and \( 1 \times 10^{-2} \), respectively.

\subsubsection*{\textbf{Multi-enhancement evaluation}}

As established in \cref{sec:findings}, multi-enhancement quality degradation is most pronounced in Case~1, where the same enhancement method is repeatedly applied.
Therefore, our evaluation focuses on this ``naive scenario'': each compressed image undergoes five enhancement cycles using the same method.
We assess the quality of enhanced images based on both fidelity and perceptual quality.
Fidelity is measured using PSNR and SSIM, while perceptual quality is evaluated using TOPIQ and LPIPS.
Additionally, we assess the degradation speed of multi-enhanced quality using the DI metric proposed in \cref{sec:findings} and the FID metric.

\subsection{Effectiveness of Domain-Consistent Quality Enhancement}

\subsubsection*{\textbf{Objective results}}

\Cref{fig:quality-degradation-exp} presents objective results, using the PSNR metric for CNN-based methods, to illustrate quality degradation under multi-enhancement across various datasets, codecs, and compression levels.
Similar trends are observed for other evaluation metrics and GAN-based methods.
Firstly, consistent with \cref{ob:1}, multi-enhancement leads to severe quality degradation for existing methods.
Secondly, our proposed method effectively reduces the degradation speed of multi-enhanced images, as demonstrated by the improved quality curves of the adapted methods compared to their baseline counterparts.
In some instances (e.g., AR-CNN~\cite{dongCompressionArtifactsReduction2015} and RDN~\cite{zhangResidualDenseNetwork2018} on both datasets), the quality degradation is slightly reduced.
In other cases (e.g., CBDNet~\cite{guoConvolutionalBlindDenoising2019} and RBQE~\cite{xingEarlyExitNot2020a} on the Flickr2K dataset), the degradation is significantly reduced, becoming almost negligible.

\begin{table}[!t]
    \caption{Average percentage difference (\%) between adapted and baseline method performance, relative to the baseline}
    \centering
    \label{tab:exp:once}

    \begin{tabular}{l c l c c c c}
        \toprule
        Codec & QP/QF & Dataset & PSNR & SSIM & LPIPS & TOPIQ \\
        \midrule
        \multirow{8}{*}{BPG}& \multirow{2}{*}{27} & DIV2K & 0.219 & 0.152 & \textbf{2.270} & 0.077\\
        & & Flickr2K & 0.160 & 0.102 & 1.843 & 0.071\\
        \cmidrule{2-7}
        & \multirow{2}{*}{32} & DIV2K & 0.166 & 0.165 & 1.587 & 0.226\\
        & & Flickr2K & 0.208 & \textbf{0.232} & 1.913 & 0.198\\
        \cmidrule{2-7}
        & \multirow{2}{*}{37} & DIV2K & 0.088 & 0.067 & 0.892 & 0.301\\
        & & Flickr2K & 0.118 & 0.048 & 0.810 & 0.299\\
        \cmidrule{2-7}
        & \multirow{2}{*}{42} & DIV2K & 0.091 & 0.206 & 1.007 & \textbf{0.394}\\
        & & Flickr2K & 0.069 & 0.218 & 0.908 & 0.364\\
        \midrule
        \multirow{8}{*}{JPEG}& \multirow{2}{*}{30} & DIV2K & 0.079 & 0.039 & 0.358 & 0.064\\
        & & Flickr2K & 0.058 & 0.049 & 0.884 & 0.065\\
        \cmidrule{2-7}
        & \multirow{2}{*}{40} & DIV2K & 0.417 & 0.192 & 1.862 & 0.219\\
        & & Flickr2K & \textbf{0.458} & 0.227 & 2.245 & 0.200\\
        \cmidrule{2-7}
        & \multirow{2}{*}{50} & DIV2K & 0.287 & 0.145 & 1.455 & 0.103\\
        & & Flickr2K & 0.293 & 0.144 & 1.592 & 0.074\\
        \cmidrule{2-7}
        & \multirow{2}{*}{60} & DIV2K & 0.212 & 0.169 & 1.351 & 0.099\\
        & & Flickr2K & 0.231 & 0.172 & 2.147 & 0.074\\
        \midrule
        \multicolumn{3}{c}{Average Percentage Difference} & 0.197 & 0.145 & 1.445 & 0.177 \\
        \bottomrule
    \end{tabular}
\end{table}

As mentioned in \cref{sec:findings}, we introduce the DI metric to measure the degradation speed of multi-enhanced quality.
A lower DI value indicates a slower degradation speed.
As shown in \cref{tab:di-improvement,tab:di-improvement-flickr2k}, our proposed method consistently adapts all baseline methods for domain-consistent quality enhancement, leading to significantly reduced DI values compared to the baseline methods.
For instance, the DI values for the adapted MPRNet method~\cite{zamirMultistageProgressiveImage2021} on the DIV2K dataset compressed by BPG are reduced by 59\% for PSNR, 46\% for SSIM, 52\% for TOPIQ, and 74\% for LPIPS.
Similar improvements are observed on the Flickr2K dataset, the JPEG codec, and other methods.
These findings indicate that our proposed method effectively reduces the multi-enhancement quality degradation speed.

Notably, the idempotency adaptation does not impair the single-enhancement performance of our adapted methods, which remains as good as that of the baseline methods.
Specifically, we measure the average percentage difference between adapted and baseline method performance, relative to the baseline.
The results are averaged over all baseline methods.
As shown in \cref{tab:exp:once}, the performance difference is less than 2.3\% across all codecs, compression levels, datasets, and metrics.

\subsubsection*{\textbf{Subjective results}}

\begin{figure}[!t]
    \centering
    \includegraphics[trim={2em 3em 2em 3em}, clip, width=\linewidth]{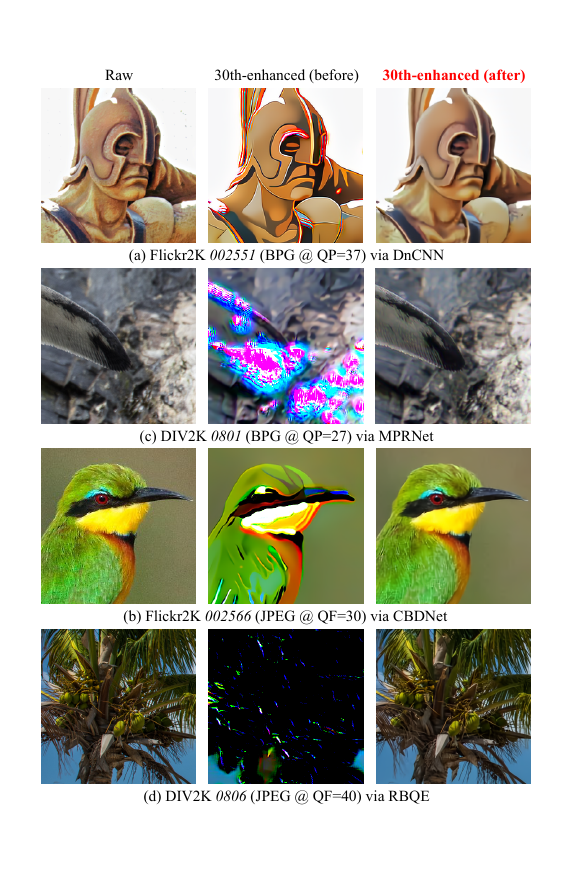}
    \caption{Subjective results illustrating quality degradation under multi-enhancement across various datasets, codecs, compression levels, and enhancement methods before and after adaptation.} 
    \label{fig:subjective-exp}
\end{figure}

As observed in \cref{ob:2}, existing methods tend to generate comic-style or oil-painting-like artifacts in the multi-enhancement scenario.
\Cref{fig:subjective-exp} presents subjective results illustrating quality degradation under multi-enhancement across various datasets, codecs, compression levels, and enhancement methods, both before and after adaptation.
We observe that baseline methods generate images with severe artifacts, such as comic-style and oil-painting-like artifacts, which are not present in the original images.
In contrast, our proposed method effectively reduces these artifacts in multi-enhanced images, resulting in more natural and visually pleasing outcomes.
More importantly, our proposed method performs well even after 30 cycles of multi-enhancement, a scenario where existing methods can generate collapsed images with severe artifacts.

\begin{figure*}[!t]
    \centering
    \includegraphics[trim={0em 0em 0em 0em}, clip, width=\linewidth]{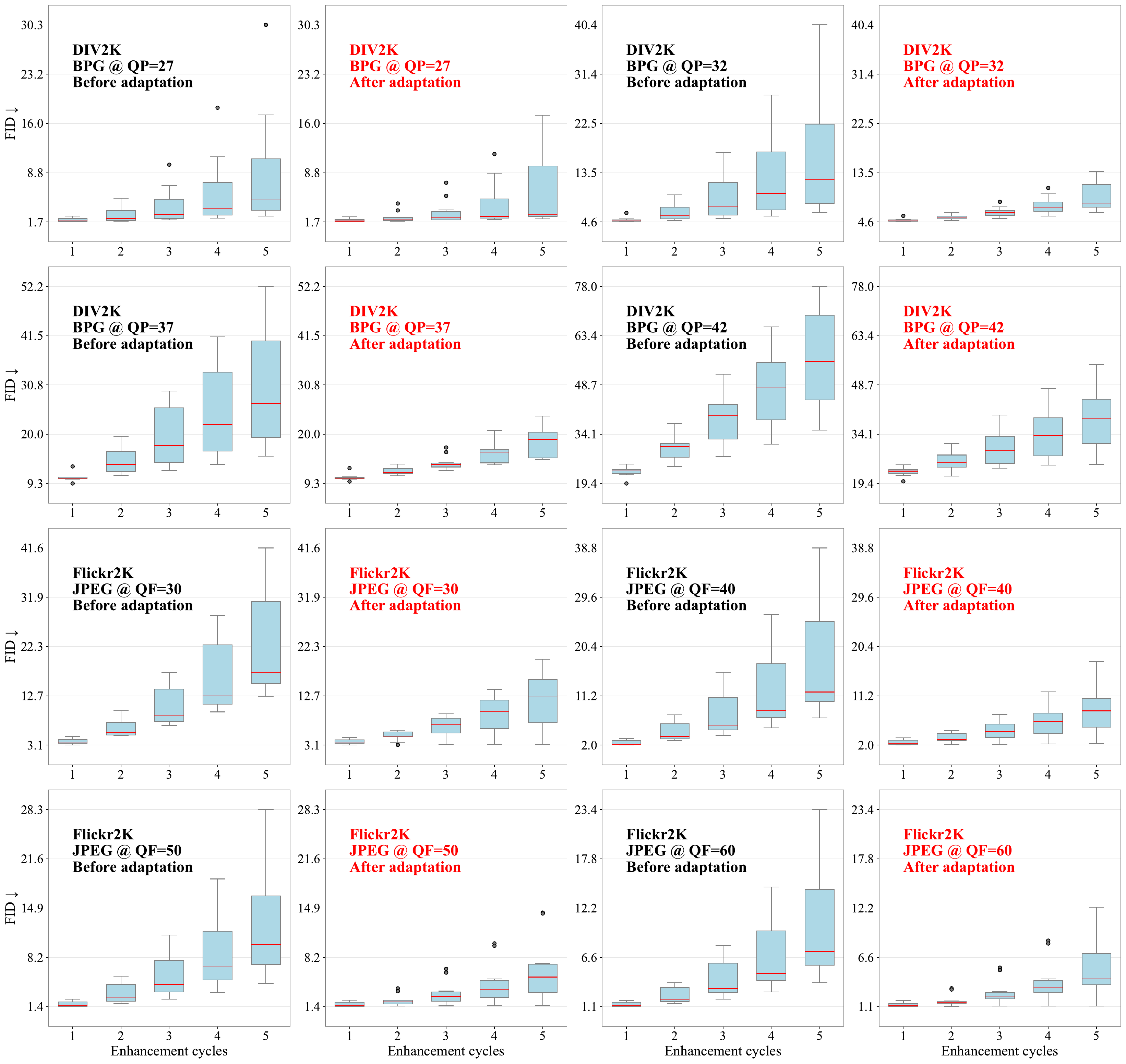}
    \caption{Objective results showing FID-based quality degradation under multi-enhancement across various datasets and codecs of all methods before and after adaptation.} 
    \label{fig:quality-degradation-fid-exp}
\end{figure*}

In addition, \cref{fig:quality-degradation-fid-exp} presents FID-based quality degradation under multi-enhancement across various datasets and codecs, comparing all methods before and after adaptation.
As shown in the figure, the adapted methods achieve significantly lower FID values than their baseline counterparts.
This indicates that our proposed method effectively reduces the domain gap between multi-enhanced images and original images, thereby improving their perceptual quality.

\section{Conclusion}

In this paper, we addressed the critical challenge of multi-enhancement robustness in quality enhancement methods for compressed images.
We demonstrated that existing enhancement methods suffer from severe degradation when repeatedly applied in real-world visual communication pipelines, generating prominent artifacts such as unnatural textures and color distortions.
To mitigate this issue, we introduced a novel domain-consistent quality enhancement framework, formulated through an idempotent enhancement method.
Our method adapts existing enhancement models by incorporating identity, idempotency, and compactness objectives, effectively maintaining image quality and domain consistency across multiple enhancement cycles.

Comprehensive experiments conducted on multiple datasets and compression settings verified the effectiveness and generalization of our method.
Results showed significant reductions in quality degradation rates under multi-enhancement scenarios, with our adapted methods outperforming existing ones both objectively and subjectively.
Future research directions include extending our idempotent training paradigm to other vision tasks and exploring adaptive mechanisms to dynamically handle diverse distortions.

\bibliographystyle{IEEEtran}
\bibliography{IEEEabrv,refs}

\appendices

\section{Proof of Theorem 1}
\label{app:1}

We define the drift measure of an instance \( x \) as:
\begin{equation}
    \dd(x) = \mathcal{D} \bigl( \func(x; \theta), x \bigr).
\end{equation}

We first show that \( \lossgroup \) minimizes the drift \( \dd \) over the target domain while maximizing it at every other point \( \func (\comp) \).  
Next, we show that \( \lossidem \) increases the probability that \( \func \) maps \( \comp \) to the low-drift region.

For simplicity, we denote \( \comp \), \( \raw \), \( \func (\cdot; \theta) \), \( \func (\cdot; \theta^*) \), \( \dnat \), \( \denh \), and \( \denhbest \) as \( z \), \( x \), \( \f(\cdot) \), \( \ff(\cdot) \), \( \px \), \( \pt \), and \( \ptb \), respectively.

We first analyze the global minimum of \( \lossgroup \) given current parameters \( \theta^* \):
\begin{equation}
    \begin{split}
        \lossgroup &= \mathbb{E}_x \bigl[ \mathcal{D} ( \f(x), x ) \bigr] - \lambda_c \mathbb{E}_z \bigl[ \mathcal{D} ( \f(\ff(z)), \ff(z) ) \bigr]\\
        &= \int \dd(x) \px(x) dx - \lambda_c \int \dd(\ff(z)) \ptb(z) dz.
    \end{split}
\end{equation}

Now, we perform a change of variables: for the left integral, let \( y := x \), and for the right, let \( y := \ff(z) \). This yields:
\begin{equation}
    \begin{split}
        \lossgroup &= \int \dd(y) \px(y) dy - \lambda_c \int \dd(y) \ptb(y) dy\\
        &= \int \dd(y) \bigl( \px(y) - \lambda_c \ptb(y) \bigr) dy.
    \end{split}
\end{equation}

Let \( M = \sup_{y_1, y_2} \mathcal{D}(y_1, y_2) \), where the supremum is taken over all possible pairs \( (y_1, y_2) \).  
Note that \( M \) may be infinite.  
Since \( \dd(y) \) is non-negative, the global minimum of \( \lossgroup \) is achieved when:
\begin{equation}
    \label{eq:lossgroup}
    \begin{split}
    \ddd(y) &= \begin{cases}
        0 & \text{if } \px(y) - \lambda_c \ptb(y) \geq 0,\\
        M & \text{otherwise}
    \end{cases}\\
    &= M \cdot \mathbb{I}_{\{ \px(y) < \lambda_c \ptb(y) \}}.
    \end{split}
\end{equation}

Next, we examine the global minimum of \( \lossidem \) given \( \theta^* \):
\begin{equation}
    \begin{split}
        \lossidem &= \mathbb{E}_z \bigl[ \mathcal{D} ( \ff(\f(z)), \f(z) ) \bigr]\\
        &= \mathbb{E}_z \bigl[ \ddd(\f(z)) \bigr].
    \end{split}
\end{equation}

Plugging in \cref{eq:lossgroup} and substituting \( \theta^* \) with \( \theta \) as we examine the inner mapping's minimum:
\begin{equation}
    \lossidem = M \cdot \mathbb{E}_z \bigl[ \mathbb{I}_{\{ \px(y) < \lambda_c \pt(y) \}} \bigr].
\end{equation}

To obtain \( \theta^* \), based on our assumption in \cref{eq:assume}, we minimize the equation:
\begin{equation}
    \theta^* = \argmin_{\theta} \mathbb{E}_z \bigl[ \mathbb{I}_{\{ \px(y) < \lambda_c \pt(y) \}} \bigr].
\end{equation}

If \( \ptb = \px \) and \( \lambda_c \leq 1 \), this reaches the minimum value of 0.  
If \( \lambda_c = 1 \), then \( \ptb = \px \) is the only minimizer, since for any \( \pt(y) < \px(y) \), there must exist some \( y' \) such that \( \pt(y') > \px(y') \), increasing the loss.  
This concludes the proof.

Additionally, \( \dd(y) \) serves as an indicator of how likely a point \( z \) is mapped \textbf{outside} the natural domain \( \px \).  
Minimizing \( \lossgroup \) helps construct an ideal indicator, while minimizing \( \lossidem \) encourages the generator to produce samples in low-energy (high-consistency) regions.
Therefore, optimizing \( \lossgroup \) and \( \lossidem \) forms an inherent adversarial generation process~\cite{goodfellowGenerativeAdversarialNets2014,shocherIdempotentGenerativeNetwork2024}, where both the ``generator'' and ``discriminator'' are optimized jointly in each iteration.
From the perspective of Energy-Based GANs (EBGANs)~\cite{zhaoEnergybasedGenerativeAdversarial2017}, \( \dd(y) \) can be interpreted as the energy that reflects idempotency between \( y \) and its mapping \( \f(y) \).

\end{document}